\documentclass[twocolumn,tighten,times,floatfix]{aastex631}
\usepackage{amsmath}
\usepackage{xspace,xcolor}
\usepackage{natbib}

\usepackage{graphicx}

\newcommand{\beq}{\begin{equation}}
\newcommand{\eeq}{\end{equation}}

\newcommand{\harm}{\texttt{HARM3D}\xspace}

\newcommand{\Bothros}{\texttt{Bothros}\xspace}

\newcommand{\Msun}{\ensuremath{M_{\odot}}\xspace}

\newcommand{\fbin}{f_{\mathrm{B}}}

\newcommand{\MdotEdd}{\dot{M}_{\mathrm{Edd}}}

\begin{document}

\title{A Parameter Study of the Electromagnetic Signatures of an Analytical Mini-Disk Model for Supermassive Binary Black Hole Systems }

\author[0000-0002-5599-6605]{Kaitlyn Porter}
\affil{Center for Computational Relativity and Gravitation, Rochester Institute of Technology, Rochester, New York 14623, USA}

\author[0000-0003-3547-8306]{Scott~C.~Noble}
\affil{Gravitational Astrophysics Lab, NASA Goddard Space Flight Center, Greenbelt, MD 20771, USA}

\author[0000-0001-7941-801X]{Eduardo~M.~Gutierrez}
\affil{Institute for Gravitation and the Cosmos, The Pennsylvania State University, University Park PA 16802, USA}
\affil{Department of Physics, The Pennsylvania State University, University Park PA 16802, USA}

\author[0000-0001-5820-8208]{Joaquin~ Pelle}
\affil{Instituto de F\'isica Enrique Gaviola, CONICET, Ciudad Universitaria, 5000 C\'ordoba, Argentina}
\affil{Facultad de Matem\'atica, Astronom\'ia, F\'isica y Computaci\'on, Universidad Nacional de C\'ordoba, Argentina}

\author[0000-0002-8659-6591
]{Manuela Campanelli}
\affil{Center for Computational Relativity and Gravitation, Rochester Institute of Technology, Rochester, New York 14623, USA}

\author[0000-0002-2942-8399]{Jeremy Schnittman}
\affil{Gravitational Astrophysics Lab, NASA Goddard Space Flight Center, Greenbelt, MD 20771, USA}

\author[0000-0002-3326-4454]{Bernard J. Kelly}
\affil{Gravitational Astrophysics Lab, NASA Goddard Space Flight Center, Greenbelt, MD 20771, USA}
\affil{Center for Space Sciences and Technology, University of Maryland
Baltimore County, 1000 Hilltop Circle Baltimore, MD 21250, USA}

\begin{abstract}
Supermassive black holes (SMBHs) are thought to be located at the centers of most galactic nuclei. When galaxies merge they form supermassive black hole binary (SMBHB) systems and these central SMBHs will also merge at later times, producing gravitational waves (GWs). Because galaxy mergers are likely gas-rich environments,  SMBHBs are also potential sources of electromagnetic (EM) radiation. The EM signatures depend on gas dynamics, orbital dynamics, and radiation processes. The gas dynamics are governed by general relativistic magnetohydrodynamics (MHD) in a time-dependent spacetime. Numerically solving the MHD equations for a time-dependent binary spacetime is computationally expensive. Therefore, it is challenging to conduct a full exploration of the parameter space of these systems and the resulting EM signatures. We have developed an analytical accretion disk model for the mini-disks of an SMBHB system and produced images and light curves using a general relativistic ray-tracing code and a superimposed harmonic binary black hole metric. This analytical model greatly reduces the time and computational resources needed to explore these systems, while incorporating some key information from simulations. We present a parameter space exploration of the SMBHB system in which we have studied the dependence of the EM signatures on the spins of the black holes (BHs), the mass ratio, the accretion rate, the viewing angle, and the initial binary separation. Additionally, we study how the commonly used fast-light approximation affects the  EM signatures and evaluate its validity in GRMHD simulations. 
\end{abstract}

\section{Introduction}

Supermassive black holes (SMBHs) are expected to reside at the center of almost all galaxies in our universe. These black holes (BHs) co-evolve with their host galaxies and over time play a major role in the formation and structure of the galaxy \citep{Heckman2014}. When galaxies merge, their SMBHs become gravitationally bound and are brought to sub-pc scales due to dynamical friction with stars and torques from the surrounding gas. We call these binary systems supermassive black hole binaries (SMBHBs). Afterward, the orbital energy loss associated with gravitational radiation will cause the orbit to shrink until they eventually merge. Merging events such as these emit powerful bursts of gravitational waves (GWs) \citep{Centrella:2010mx, Burke-Spolaor2019}.

The frequency at which GWs are emitted scales with the inverse of the total mass of the binary system ($f \propto M^{-1}$). Current ground-based GW observatories, like the Laser Interferometer Gravitational Wave Observatory (LIGO) and the VIRGO interferometer, have detected stellar-mass binary black hole systems \citep{LIGOScientific:2016aoc, LIGOScientific:2020ibl,LIGOScientific:2021usb, KAGRA:2021vkt}. Ground-based observatories can detect GW frequencies on the order of  $10^1 - 10^3 $ Hz. These observatories are limited by the fact that Earth's seismic activity can interfere with detections at lower frequencies.

Because the total mass of an SMBHB system is much greater than a stellar-mass binary system, the GW frequency that these supermassive systems emit will fall outside of LIGO and VIRGO's detection range. Space-based observatories such as the Laser Interferometer Space Antenna (LISA), an ESA-NASA mission with plans to launch in the mid-2030s, are therefore necessary to detect their GWs \citep{LISARedbook}. 

SMBHBs are also expected to be detectable by pulsar timing arrays (PTAs, \cite{Burke-Spolaor2019}). Currently, there are five PTA projects/collaborations that are active: the Parkes Pulsar Timing Array (PPTA, \cite{Hobbs:2017oam}), European Pulsar Timing Array (EPTA, \cite{Babak2016}), North American Nanohertz Observatory for Gravitational Waves (NANOGrav, \cite{NANOGrav:2023gor}), Five-hundred-meter Aperture Spherical Radio Telescope (FAST, \cite{Hobbs2019}), and Indian Pulsar Timing Array (InPTA, \cite{Tarafdar2022}). PTAs are expected to be able to detect binary black hole systems with total masses around $10^{8} - 10^{10} \Msun$. It is expected that detection with PTAs of an individual SMBHB system could occur within the next few years and that detection rates should increase steadily as PTAs improve through longer observational times and with the inclusion of more pulsars \citep{BogdanovicReview}. 

Less massive systems of the range $10^3 - 10^7 \Msun$ should be detectable by LISA. LISA is expected to detect at least a few SMBHB mergers per year \citep{LISARedbook,Klein2016,DalCanton2019}.

Because merging galaxies are gas-rich environments, there is likely to be matter surrounding the binary system as well as accreting around the individual BHs \citep{hopkins10,Chapon2013,Pfister2017,Tremmel2017, Gutierrez_etal2024a}. Therefore, SMBHBs are also potential sources of EM  radiation. Due to the complexity of these systems, the nature of this radiation is still uncertain.  A major question that needs to be addressed is how we can distinguish between single SMBHs at the center of galaxies and binary black hole systems using the EM radiation coming from these sources. Because the spatial scales are unresolvable with current GW instruments, these binaries need to be identified by indirect means, through their potential distinctive EM signatures. Understanding the structure of the accretion flow onto SMBHBs is crucial for predicting the possible EM emissions associated with the accreting matter, and determining the relevant emission sites, (quasi-)periodicities, and spectral properties. Modeling these processes is a very challenging task, involving magnetohydrodynamic (MHD) simulations of the accretion disks around these systems with dynamical spacetimes, and the radiative transfer of the resulting EM emission to predict the observable features (see \citealt{Gutierrez_etal2024a} for a review of the topic).

During the relativistic regime, when gravitational radiation controls the binary's orbital evolution, a circumbinary disk (CBD) surrounds the system, and matter is accreted into the binary in the form of two narrow streams originating from the inner edge of the CBD \citep{Noble2012}. These streams feed a pair of accretion disks around each of the BHs, called ``mini-disks". Additionally, it has been found that matter traveling along these streams can be redirected back toward the CBD, creating an overdensity known as the ``lump'' at the inner edge of the CBD \citep{Shi12, Noble2012}. 

The accretion of matter onto the mini-disks is modulated by these lumps. When one of the mini-disks travels close to the lump it can pull mass from the lump along the accretion streams. For near equal-mass binaries, this occurs at twice the beat frequency,  $f_{\mathrm{beat}}=\fbin-f_{\rm lump}$, which is the difference between the orbital frequency of the binary and the orbital frequency of the lump \citep{Noble2012,Bowen2018,Bowen2019}. 

If the accretion onto the BHs varies due to this overdensity then there would be some variability in the spectra of the mini-disks as well, which would most likely be detectable in the X-ray frequencies \citep{D'Ascoli2018, Gutierrez:2021png}. It has also been found that the amplitude of the lump depends on the magnetization of the accretion material \citep{Noble:2021vfg}. The greater the magnetic field the less dense the lump becomes, consequently weakening the EM signal as well. The strength of the lump and the resulting EM radiation also depends on the mass ratio of the binary system ($q \equiv M_{2}/M_{1}$, where $M_{1}$ is the mass of the primary BH and $M_{2}$ is the mass of the secondary BH). Both the lump and the EM radiation decrease with smaller mass ratios, with the lump feature vanishing entirely for $q$ somewhere between 0.5 and 0.2 \citep{Noble:2021vfg}.

For binaries at close separations, a mass exchange process has been found to occur between the two mini-disks. This ``sloshing'' effect of the gas occurs at a frequency of around 2-3 times $\fbin$, where $\fbin$ is the binary orbital frequency, and can cause modulations in the mass of the mini-disks which would show up as variability in the EM signatures \citep{Bowen:2016mci, Avara2023}.

EM emission is expected to be produced in the CBD, mini-disks, and accretion streams. The phenomenology associated with this depends on many parameters, such as the mass accretion rate, the binary’s total mass $M$, the mass ratio $q$ of the BHs, the viewing angle, and the spins of each of the BHs. The first EM signatures of SMBHBs using general relativistic magnetohydrodynamic (GRMHD) simulations of the gas surrounding the system were modeled by \citet{D'Ascoli2018}. The time-varying spectra were studied for non-spinning binary systems with a total mass of $10^{6} \Msun$ and they found that the mini-disks surrounding each of the BHs were the most energetic features in the system. These mini-disks emit light along the visible to X-ray wavelengths, with their luminosity peaking around $10^{16}$ Hz ($\sim 40$ eV).
In 2022, \citet{Gutierrez:2021png} modeled the light curves and spectra of equal-mass SMBHBs for both spinning and non-spinning systems using GRMHD simulated data \citep{Bowen17, Bowen18, Combi:2021xeh}. They explored the dependence of the spectra on the total mass of the system and found that the luminosity of the system increases as the total mass of the BHs increases. Additionally, it was found that the EM signatures show periodicity associated with the lump modulating both the accretion rate and the mass of the mini-disks. The characteristic frequencies of this variation are $f\sim 0.2 \fbin$ and $f\sim 1.4 \fbin = 2 f_{\rm {beat}}$. The former is the frequency of the radial oscillation of the lump since it has a slightly eccentric orbit and the latter is the frequency of the matter accreting onto one of the mini-disks from the lump. 

\citet{Gutierrez:2021png} also showed that in the case of spinning BHs with spin parameters of $0.6$, the mini-disks are on average three to five times brighter compared to those of a non-spinning binary system. In the resulting spectra the luminosity increases in the far-UV and soft X-ray bands. Additionally, at later times in the inspiral, the mini-disks do not contribute to the total spectrum of the system for non-spinning BHs, but the mini-disks around spinning BHs do contribute to the spectrum. This is attributed to the fact that the accretion disks around spinning BHs are more massive. Gas around a spinning BH can maintain circular orbits closer to the event horizon compared to a non-spinning BH.

In addition to the accretion structures already discussed, there is also expected to be a corona composed of ionized gas and scattered photons around each mini-disk. Because of this, the X-ray spectra of SMBHB systems could be of particular interest due to the possible presence of the iron (Fe) K$\alpha$ fluorescent emission line, a prominent component of the X-ray reprocessed disk spectra. The iron line has been extensively studied and observed for single-BH systems \citep{Porter2020, Fabian1989,Young2000,Reynolds1999,Tanaka1995,Wilkins2021}. There has also been some progress made in simulating the reprocessed spectra around binary systems, and some characteristic spectral features have been found that arise in the Fe K$\alpha$ emission line that, if observed, might indicate the presence of a binary system \citep{Yu2001,Sesana2012,McKernan2013,Javanovic2014}.

In addition to EM signatures arising from processes occurring within the accretion flows around the binary system, there are also interesting signatures that could arise from the orbital motion of the BHs. When viewing these systems close to edge-on, it is likely that the binary system will self-lens, meaning that one of the BHs will act as a gravitational lens while the other acts as a light source. This self-lensing should cause periodic flares in the light observed from these systems that occur at twice the binary's orbital frequency \citep{SLF00,Schnittman:2018,SLF01,SLF1,SLF2,SLF3}.

Once sufficiently distinctive EM signatures are defined, this knowledge can be used to guide future wide-field survey telescopes looking for potential binary systems such as SDSS-V (\cite{SDSS2019}), eROSITA (\cite{eROSITA2021}), and in the near future the Vera Rubin Observatory (\cite{LSST2019}). The possible candidates can be further refined through narrow-field telescope missions like the Hubble Space Telescope (HST) and Chandra. Post-merger systems might also be targets for the James Webb Space Telescope (JWST) \citep{Schnittman2008}. Although GW detections have yet to be made of these systems, if EM observations are made before a GW detection it will improve the efficacy of GW signal searches with LISA and PTAs considerably, as well as significantly reduce the uncertainty in estimates of the potential LISA source population. Complementary EM and GW data will enable a more complete understanding of the immediate environments close to SMBHB mergers.

GRMHD simulations of the accretion structures surrounding a binary black hole system are computationally very expensive. The goal of this paper is to present an analytical thin accretion disk model of the mini-disks surrounding each BH. An analytical disk model incorporated into the framework of the ray-tracing code, \Bothros, greatly reduces the computational time needed to calculate the EM signatures of these systems. By fitting this model as close as possible to MHD simulations, we are able to explore these systems more thoroughly and see how the images and light curves from these systems are affected by changing key parameters such as the total mass of the system, the spins of the BHs, and the mass ratio. Additionally, by saving computation time with an analytical disk model, we are able to explore the effects of using the fast light approximation (FLA) on the EM signatures. This approximation is often used when ray-tracing with GRMHD data and therefore it is important to test its accuracy for future GRMHD ray-tracing simulations.

This paper is organized as follows.  In Section \ref{sec:model}, we discuss our semi-analytical metric, ray-tracing code, and analytical accretion disk model.  Then in Section \ref{sec:model}, we introduce the parameter space exploration we conducted and present our results and their implications. Then we summarize and conclude our findings in Section \ref{sec:discussion}.

\section{Model}\label{sec:model}
\subsection{Superposed-PN Metric}

 For our binary spacetime, we utilize the superposed post-Newtonian (SPN) metric developed by
 \citet{Combi:2021xeh}. The need for this type of semi-analytical spacetime metric arises from the fact that numerical spacetime metrics place a limit on the time GRMHD simulations of binary systems can run. Numerical spacetimes are computationally demanding to calculate and thus the simulations that run for the length of time required for the system to reach a steady state are currently too expensive. The SPN metric's computational efficiency is ideal for a parameter space study, in which we want to run multiple simulations. We have also chosen to use this metric because it allows for non-zero-spin binary systems and it has already been implemented and well-tested within the radiative transfer code, \Bothros \citep{ Gutierrez:2021png}.  MHD simulations of the CBD and mini-disks using \harm have already been run using this metric \citep{Combi:2021dks} and within \Bothros it has been used to produce light curves and spectra from GRMHD data of SMBHB systems \citep{Gutierrez:2021png}. The SPN metric is a superposition of two BH metrics in harmonic coordinates and it takes the form of 

\begin{align}\label{eq:KSg1}
     g_{\mu\nu}^{PN}= \eta^{H}_{\mu\nu} + M_{(1)} \mathcal{H}^{(1)}_{\mu\nu} + M_{(2)} \mathcal{H}^{(2)}_{\mu\nu},
\end{align}

where $M_{(1)}$ and $M_{(2)}$ are the masses for each BH and  $\mathcal{H}^{(1)}_{\mu\nu}$ and $\mathcal{H}^{(2)}_{\mu\nu}$ are tensor terms for each of the BHs which encode information about their mass and spin and the coordinate system being used. The tensor terms also include a boost transformation and a transformation from local Kerr--Schild coordinates to local Cook--Schild harmonic coordinates, with the latter transformation allowing us to use higher-order post-Newtonian (PN) accurate trajectories.  The instantaneous boosts are required to include the motion of the black holes in a covariant way.  Note that this spacetime, including the boost,  was found to produce insignificant violations of Einstein's equations far from the black holes \citep{Combi:2021xeh}. 

In order to describe the binary system over time, the metric must be supplied with the position, velocity, and acceleration vectors of each of the BHs. To supply this information, the orbital phase $\Phi(t)$ and binary separation $r_{12}(t)$ are obtained by solving the post-Newtonian (PN)  equations of motion to 3.5 PN order.

In general, PN theory is a good approximation when considering a compact binary system at large separations. This SPN metric is restricted to BH separations of $10M$ or larger. For separations smaller than $10M$ the SPN metric breaks down. However, it can still be used to understand the spacetime structure and null-geodesic paths close to the event horizon because the PN approximation is only used to calculate the trajectories of the BH. The metric itself is the superposition of two boosted black hole metrics which accurately describes the spacetime in their vicinity. 

\subsection{Ray-Tracing}
To ray-trace light around our binary black hole model we are using the general-relativistic ray tracing code \Bothros. \Bothros solves both the geodesic and radiative transfer equations to produce time and frequency-dependent images, light curves, and spectra. Previously, \Bothros has been used to calculate the EM signatures around single black holes and binary black hole systems using simulated GRMHD data \citep{Noble2009a, Noble2009b, D'Ascoli2018, Gutierrez:2021png}.

\Bothros uses a camera-to-source method in integrating the geodesic equations. In this approach, the light rays start at a fixed location chosen for the camera and \Bothros integrates backward in time to the source. Tracing the photons from the camera to the source is computationally practical in that it allows us to only track the photons arriving at the distant observer.

Previously, \Bothros solved the geodesic equations under the fast light approximation, though see \citet{Noble2009b} when the approximation was dropped to explore its effect on single-BH coronal emission variability. Under this approximation, the light travel time across the source is assumed to be negligible compared to the dynamic timescale of the spacetime. Therefore, for each single snapshot, the positions of the BHs are held fixed while the photon geodesics are solved. The FLA drastically reduces the computational cost of radiative transfer, especially when ray-tracing data from GRMHD simulations, where interpolation is the most important bottleneck. However, since we consider an analytical accretion disk model, we can drop this approximation without significant overhead. We take advantage of this to test the range of validity of the approximation by comparing results with and without it for some of the runs in our parameter space exploration.

Since \Bothros mimics an observer located at a large distance from the binary system, the photons travel over a large amount of empty space. In order to do this efficiently, the size of the integration step is large when traveling through flatter regions of space and is adaptively made smaller when the ray travels closer to the BHs and gravitational effects start to more significantly impact the path of the photons.

When \Bothros calculates the path of light rays, there are a few instances when geodesic integration should be stopped. The first instance would be when a photon encounters the event horizon of one of the BHs. For the geodesic calculation here we are defining the event horizon to be the outer event horizon of a Kerr BH in Boyer--Lindquist coordinates. Any photon that reaches  within a buffer value of $1M_{i}$ from the horizon radius, $r_{+,i}^{\rm BL} = M_i + \sqrt{M_i^2 - a_i^2}$, is specified as residing ``within the event horizon." Since the integration is backward in time, this would correspond to a photon originating from just outside the event horizon. When this occurs, the intensity of the photon is initialized to zero.

Another instance at which the integration is stopped is if the photon starts to diverge away from the binary system. This would correspond to photons that are originating outside of the binary system, which we do not want to consider. Again, the intensity in this instance is set to zero. Lastly, integration is also stopped when the geodesic path encounters the surface of the accretion disk. Because our disk model is geometrically thin and optically thick, we do not expect to receive any photons coming from any depth below the surface of the disk. When a photon encounters the disk surface, the integration is stopped and the intensity of the light at that location is calculated using the analytical thin disk model. 

For our parameter space exploration, the resolution of the ``camera" in \Bothros has been set to $1000\times 1000$ pixels for all runs. The solid angle covered by the camera observing a spherical source of radius $d$ is $\Omega = 2\pi (1 - r_{\rm{cam}}/\sqrt{r_{\rm{cam}}^2 + d^2})$ where $r_{\rm{cam}}$ is the distance from the camera to the center of mass of the binary system. For all of the following runs we have set $r_{\rm{cam}} = 1000M$ and $d$ varies depending on the initial separation and mass ratio. The chosen resolution of 1000x1000 is a converged resolution such that the observed flux for a full orbit of the binary system changes by less than 1\% when compared to higher resolutions. In the ray-tracing scheme used, the geodesic equations are written in the form of eight first-order equations representing the photons' positions and velocities over time. The time component of the four-velocity equation can be eliminated by solving for the null velocity normalization condition, $u^{\mu}u_{\mu} = 0$, thus reducing the number of geodesic equations that \Bothros needs to solve down to seven. 

\subsection{Radiative Transfer}

After the path of every photon is calculated for a single snapshot, the photons that end at the surface of the accretion disk are given an intensity value that is determined through the analytical accretion disk model. With the initial intensity at the disk calculated, \Bothros then solves the radiative transfer equation forward in time to find the intensity of the light at the camera. This calculation takes place after the geodesic calculation and it uses the same path and step sizes found in that integration. 

The covariant radiative transfer equation is:

\begin{align}
    \frac{\mathop{d}}{\mathop{d\lambda}} \left( \frac{I_\nu}{\nu^3}\right) &= \frac{j_\nu}{\nu^2} - \nu \alpha_\nu \left( \frac{I_\nu}{\nu^3}\right) \,,
    \label{eq:transfer_equation}
\end{align}

where $\lambda$ is an affine parameter, $I_\nu$ is the specific intensity of the radiation field at frequency $\nu$, and $j_\nu$ and $\alpha_\nu$ are the plasma emissivity and absorptivity coefficients, respectively. Each term in the equation is separately Lorentz invariant. For our model, we set both $j_\nu$ and $\alpha_\nu$ to zero, meaning we assume that no emission or absorption processes occur between the source and the observer. This implies that the invariant quotient $I_\nu/\nu^3$ does not change along the geodesics. However, the specific intensity at the disk and at the observer is not the same due to the effects of redshifting or blueshifting of the frequency of the light.

With the specific intensity calculated for each pixel at a given time, we can calculate the observed flux at that frequency by integrating the specific intensity of every pixel in the frame over coordinates at the camera ($x$, $y$).

\begin{align}
    F_{\nu} = \int I_{\nu} dx dy
\end{align}

Then the specific luminosity at the frequency $\nu$ is calculated as

\begin{align}
    L_{\nu} = F_{\nu} 4 \pi r_{\rm{cam}}^2
\end{align}

\subsection{Analytical Mini-Disk Model}
\subsubsection{Novikov--Thorne Accretion Disk}

One of the simplest approaches to a thin accretion disk model around a BH is the Novikov--Thorne (NT) accretion disk \citep{Novikov:1973, 1974Page&Thorne}. The NT model makes several assumptions resulting in an analytical expression for the local flux from the disk. Firstly, the disk and BH spacetime are assumed to be axisymmetric and stationary, with the disk having negligible self-gravity. NT disks are also assumed to be perfectly radiatively efficient, meaning any heat generated from work done by internal stresses is assumed to be radiated away locally and instantaneously. They are also often assumed to have infinite optical depth. With these assumptions, one can show that NT disks are geometrically infinitesimally thin and lie along the plane orthogonal to the BH's spin axis. The disk extends only to the innermost stable circular orbit (ISCO), where one assumes the internal stress becomes zero. This means that no light is emitted within the ISCO from NT disks. Although we are modeling an inspiraling binary system, this stationary requirement is met by the fact that all flux calculations are done in each of the BH's rest frames, in which the BHs are stationary. 

In our model of the mini-disks, we set the accretion disks to lie on the orbital plane of the binary. To construct a mini-disk around each BH based on the NT model, it is necessary to connect the global coordinates with the Boyer--Lindquist coordinates of each BH. To do this, we first transform to the Kerr--Schild coordinates relative to the corresponding BH, including the translation and boost described in \citet{Combi:2021dks}, and then perform the usual transformation to Boyer--Lindquist coordinates.

In these coordinates, we assume that each mini-disk extends between $r_{\text{isco}} \leq r \leq r_{\text{trunc}}$, where $r_{\text{isco}}$ is the ISCO radius of the Kerr spacetime with the mass and spin of the respective BH \citep{1974Page&Thorne}. On the other hand, we set the truncation radius of the mini-disks at $r_{\text{trunc}} = 0.4 r_{12}$, where $r_{12}$ is the binary separation (\cite{Bowen2017}). 

Next, we define the physical quantities of the model following the prescriptions for a single BH. We assume that the particles in each disk follow pseudo-Keplerian orbits, where

\begin{align}
    \Omega &= \frac{d\phi}{dt} = \frac{-g_{t\phi, r} + \sqrt{(g_{t\phi, r})^2 - g_{tt, r} g_{\phi\phi, r}}}{g_{\phi\phi, r}}\,,\\
    \tilde{E} &= - \frac{g_{tt}+ g_{t \phi} \Omega}{\sqrt{-g_{tt} - 2g_{t\phi}\Omega - g_{\phi\phi} \Omega^2}}\,,\\
    \tilde{L} &= \frac{g_{t\phi}+ g_{\phi \phi} \Omega}{\sqrt{-g_{tt} - 2g_{t\phi}\Omega - g_{\phi\phi} \Omega^2}}
\end{align}

are the angular velocity, the specific energy, and the specific angular momentum of the circular orbits, respectively. The four-velocity of the gas is first calculated as in a single BH:

\begin{align}
    u^t &= \frac{g_{t \phi} \tilde{L}  + g_{\phi \phi} \tilde{E}}{g^2_{t \phi} -g_{tt}g_{\phi\phi}} \label{eq:4velcirc1}\\
    u^r &= 0 \label{eq:4velcirc2}\\
    u^{\theta} &= 0 \label{eq:4velcirc3} \\
    u^{\phi} &= \frac{g_{t t} \tilde{L}  + g_{t \phi}\tilde{E}}{g^2_{t \phi} -g_{tt}g_{\phi\phi}} \label{eq:4velcirc4}
\end{align}

and then it is renormalized so that $ g^{PN}_{\mu\nu} u^\mu u^\nu = -1 $, where $ g^{PN}_{\mu\nu} $ is the full binary BH metric of \eqref{eq:KSg1}. This four-velocity determines the local rest frame of the mini-disk. In this frame, we assume that the emitted flux as a function of radius is given by:

\begin{equation}
    F(r) = \frac{\dot m c^2}{4 \pi r^2} f(x)
\end{equation}

\begin{equation}
    \begin{split} 
        f(x) = \frac{3}{2} \frac{1}{x^3 -3x +2\chi}
        \biggr[ x- x_{0} - \frac{3}{2} \chi \ln \left(\frac{x}{x_{0}}\right)\\ - \frac{3(x_{1} - \chi)^2}{x_1(x_1 -x_2)(x_1-x_3)}\ln\left(\frac{x-x_1}{x_0-x_1}\right)\\
        - \frac{3(x_2 - \chi)^2}{x_2(x_2 - x_1)(x_2 - x_3)} \ln\left(\frac{x-x_2}{x_0-x_2}\right)\\
         - \frac{3(x_3 - \chi)^2}{x_3(x_3 - x_1)(x_3 - x_2)} \ln \left(\frac{x-x_3}{x_0-x_3}\right) \biggr] 
    \end{split} 
\end{equation}

\begin{align*}
    x &= \sqrt{r / r_g}\,, \\
    x_0 &= \sqrt{r_{\text{isco}}/r_g}\,,\\
    x_1 &= 2 \cos\left(\frac{1}{3} \arccos (\chi) - \frac{\pi}{3}\right)\,, \\
    x_2 &= 2 \cos\left(\frac{1}{3} \arccos (\chi) + \frac{\pi}{3}\right)\,, \\
    x_3 &= -2 \cos\left(\frac{1}{3} \arccos(\chi)\right)\,.
\end{align*}

In the equations above, $\dot{m_i}$ is the mass accretion rate of the BH  where $i$ specifies the particular BH around which these values are being calculated, $\chi_{i} = a_{i}/M_{i}$ is the dimensionless spin parameter of the BH, and $r_g = M_{i}$ is the gravitational radius of the BH \citep{1974Page&Thorne}. 

Additionally, we assume the disk emits like a blackbody in the local rest frame so that the local effective temperature is given by the Stefan-Boltzmann Law:
\begin{equation}
\label{eq:T}
    T(r) = (F(r)/\sigma)^{1/4},
\end{equation}
where $\sigma$ is the Stefan-Boltzmann constant. The specific intensity at the disk can be calculated as:
\begin{equation}\label{eq:Inu}
     I_{\nu} = \frac{(2h/c^2)\nu^3}{e^{h\nu/kT} - 1} \, .
\end{equation}

\subsubsection{Smoothly Broken Power-Law Profile}

GRMHD simulations have shown that there is a significant amount of flux originating from within the ISCO because internal stresses are not expected to completely vanish within this radius \citep{Noble2009a,Noble:2010mm}. This emission within the ISCO results in an increase in flux at the higher energy end of the spectrum and the NT disk models do not fit this spectral shape \citep{Noble:2011wa}. Therefore, in addition to the Novikov--Thorne analytical model, we also implemented a smoothly broken power law (SBPL) analytical emission model for both mini-disks, which was introduced by \citet{Schnittman2016} for a single spinning BH.

The SBPL model that \cite{Schnittman2016} developed includes radiating material within the ISCO to better predict the spectrum of an accretion disk around a single spinning BH. They compared their GRMHD simulations of a single BH accretion disk using \harm of varying spins $(a = 0.0,0.5,0.9,0.99)$ to the NT and SBPL models with the same spin parameters. They found that the observed spectra from the simulations consistently correspond to an NT model with a higher spin, whereas the SBPL models closely match the correct spin.

The location of the innermost stable circular orbit depends on the BH spin. To compare the MHD simulations to each other, as well as to the analytical model, \cite{Schnittman2016} first rescale the radial coordinate. The following transformation was used from the radial coordinate in Boyer--Lindquist coordinates, $r$, to a new radial coordinate, $\tilde{r}$:
\begin{equation}
    d\tilde{r} = g_{rr}^{1/2} dr = \left(1 - \frac{2}{\rho} + \frac{\chi^2}{\rho^2} \right)^{-1/2} dr \, ,
\end{equation}
where $\rho = r/M_i$ is the dimensionless radial variable. Solving this integral yields
\begin{align}
    \tilde{r} = \tanh^{-1}\left(\frac{\rho-1}{\sqrt{\chi^2 - 2\rho + \rho^2}}\right) + \sqrt{\chi^2 -2\rho + \rho^2} + k
\end{align}
where the constant of integration $k$ can be chosen such that $\tilde{r}=r$ at a particular $r$. If this condition is true when $r = 3 r_{\rm{isco}}/2$ and we define the dimensionless ISCO radius to be $\rho_{\rm{isco}} = r_{\rm{isco}}/M_i$, then the integration constant is:

\begin{align}
\begin{split}
    k = \frac{3}{2}\rho_{\rm{isco}} - \tanh^{-1}\left(\frac{\frac{3}{2}\rho_{\rm{isco}} -1}{\sqrt{a^2 + (\frac{3}{2}\rho_{\rm{isco}} - 2)\frac{3}{2}\rho_{\rm{isco}}}}\right) \\ - \sqrt{\chi^2 + (\frac{3}{2} \rho_{\rm{isco}} -2 )\frac{3}{2}\rho_{\rm{isco}}} \ .
\end{split}
\end{align}

Lastly, in order to make the new radial coordinate dimensionless another re-scaling is done, where $r^*\;\equiv \;\tilde{r}/{r}_{{\rm{isco}}}$:

\begin{align}
    r^{*} = \frac{\rho}{\rho_{\text{isco}}} - \frac{2\ln(\rho)}{\rho_{\text{isco}}} - \frac{\chi^2}{\rho\rho_{\text{isco}}} + 2\ln(\frac{3\rho_{\text{isco}}}{2}) + \frac{2\chi^2}{2\rho_{\text{isco}}^2}
    \ .
\end{align}

In terms of this new radial coordinate, the new analytical fit takes the form of a smoothly broken power law:
\begin{align}
    \frac{dL}{dr^*} = C r^{* \varphi} \biggr[\frac{\cosh(\ln(r^{*}/R_{0})/\Delta R)}{\cosh(\ln(1/R_{0})/\Delta R)}\biggr]^{\xi \Delta R} \ , 
\end{align}
where $C$ is a normalization constant, $R_0$ is the location of the break in the power law, and $\Delta R$ is the width of the break. The slope of the power law is given by $\alpha$ and $\beta$ where $\varphi = (\beta + \alpha)/2$ and  $\xi = (\beta - \alpha)/2$. The free parameters that produced the best fit with the GRMHD simulations have the following values: $\beta = -2$, $\alpha = 1.73$, $R_{0} = 1.68$, and $\Delta R = 0.92$.

The normalization constant, $C$, is a function of the disk's luminosity, $L$, and is proportional to the mass accretion rate $\dot m$. This can be found by normalizing the SBPL flux to the NT flux at large distances from the BHs. For this model, we calculated the normalization constant by solving for both observed fluxes from the disks at a location of $r = r_{\text{cam}} = 1000M$:

\begin{align}
    C = \frac{F_{\text{NT}}(r_{\text{cam}})}{F_{\text{SBPL}}(r_{\text{cam}})}
\end{align}

The flux at a radius $r$ on the disk is given by: 

\begin{align}
    F(r^{*}) = \frac{1}{4 \pi r^{*}} \frac{dr^*}{dr} \frac{dL}{dr^*}
\end{align}

As with the NT model, the local temperature and intensity are calculated with equations (\ref{eq:T}) and (\ref{eq:Inu}). The advantage of the SBPL model is that it allows for the presence of gas within the ISCO. Outside of the ISCO, we keep the same four-velocity as given in equations (\ref{eq:4velcirc1}-\ref{eq:4velcirc4}). The gas within the ISCO is no longer following quasi-Keplerian orbits, so we must also define a new four-velocity profile:

\begin{eqnarray}
u^t &=& -\tilde{E} g^{tt} + \tilde{L} g^{t \phi}\\
u^r &=& u_r g^{rr}\\
u^{\theta} &=& 0\\
u^{\phi} &=& - \tilde{E} g^{t \phi} + \tilde{L} g^{\phi \phi} \ ,
\end{eqnarray}
with the covariant radial component expressed as:
\begin{align}
    u_{r} = - \sqrt{\frac{1 + \tilde{E}^2 g^{tt} - 2 \tilde{E}\tilde{L} g^{t \phi}}{g^{rr}}} \ .
\end{align}

\section{Results} \label{sec:results}

\subsection{Exploration of the Parameter Space}

Table \ref{table:PSE} describes the parameter space explored. For this study, we will look at how the initial separation of the BHs $r_{12}$, the mass ratio $q = M_2/M_1$, the accretion rate $\dot{M} = \dot{m_i}/\dot{M}_{\text{Edd}}$, the spin parameter $a_{i}$, and the viewing angle $\theta$ all affect the resulting light curves of this analytical mini-disk model. For each of the models shown in Table \ref{table:PSE}, we will be using the SBPL analytical mini-disk model. The fiducial model is M0. We will be comparing all the models to this reference model. We will also be testing how the fast light approximation affects the light curves. For each parameter that we change, we will run with and without the fast-light approximation. \Bothros calculates both time- and frequency-dependent images. All runs have a frequency range of $10^{15}$ to $10^{18}$ Hz. In the following results section, we have chosen to show the light curves at a frequency of $1.58 \times 10^{16}$ Hz because the mini-disks of a  $10^6 \Msun$ binary system emit across visible and X-ray frequencies, but their spectra peak around $10^{16}$ Hz. 

\begin{table}[ht]
\centering
\begin{tabular}{ |c||c|c|c| c | c | c| c| }
\hline
 Model ID & FLA & $a$ & $r_{12}$ initial & $r_{12}$ final & $q$ & $\theta$ & $\dot{M}$ \\
 \hline
 \multicolumn{8}{|c|}{Reference Model} \\
 \hline
 M0 & Yes  & 0.0 & 30M & 10M & 1.0 & 90$^{\circ}$ & 0.5\\
 \hline
 M1 & No & 0.0 & 30M & 10M & 1.0& 90$^{\circ}$ & 0.5\\
 \hline
 \multicolumn{8}{|c|}{Black Hole Separation Dependence} \\
 \hline
 M2 & Yes  & 0.0 & 50M & 20M & 1.0 & 90$^{\circ}$ & 0.5\\
 \hline
 M3 & No & 0.0 & 50M & 20M & 1.0 & 90$^{\circ}$  & 0.5\\
 \hline
  \multicolumn{8}{|c|}{Accretion Rate Dependence} \\
 \hline
M4 & Yes & 0.0 & 30M & 10M & 1.0 & 90$^{\circ}$  & 0.1\\
 \hline
  \multicolumn{8}{|c|}{Mass Ratio Dependence} \\
 \hline
M5 & Yes  & 0.0 & 30M & 10M & 0.50 & 90$^{\circ}$ & 0.5\\
 \hline
M6 & No & 0.0 & 30M & 10M & 0.50 & 90$^{\circ}$ & 0.5\\
 \hline
M7 & Yes & 0.0 & 30M & 10M & 0.20 & 90$^{\circ}$ & 0.5\\
 \hline
M8 & No & 0.0 & 30M & 10M & 0.20 & 90$^{\circ}$ & 0.5\\
 \hline
M9 & Yes & 0.0 & 30M & 10M & 0.10 & 90$^{\circ}$ & 0.5\\
 \hline
M10 & No & 0.0 & 30M & 10M & 0.10 & 90$^{\circ}$ & 0.5\\
 \hline
  \multicolumn{8}{|c|}{Spin Dependence} \\
 \hline
M11 & Yes & 0.3 & 30M & 10M & 1.0 & 90$^{\circ}$ & 0.5\\
 \hline
M12 & No & 0.3 & 30M & 10M & 1.0 & 90$^{\circ}$ & 0.5\\
 \hline
M13 & Yes & 0.6 & 30M & 10M & 1.0 & 90$^{\circ}$ & 0.5\\
 \hline
M14 & No & 0.6 & 30M & 10M & 1.0 & 90$^{\circ}$ & 0.5\\
 \hline
 \multicolumn{8}{|c|}{Viewing Angle Dependence} \\
 \hline
M15 & Yes & 0.0 & 30M & 10M & 1.0 & 0$^{\circ}$ & 0.5\\
 \hline
M16 & Yes & 0.0 & 30M & 10M & 1.0 & 45$^{\circ}$ & 0.5\\
\hline
\end{tabular}
\caption{Parameter Space Exploration}
\label{table:PSE}
\end{table}

\subsection{Binary Gravitational Lensing}

\begin{figure*}[h!]
\gridline{\fig{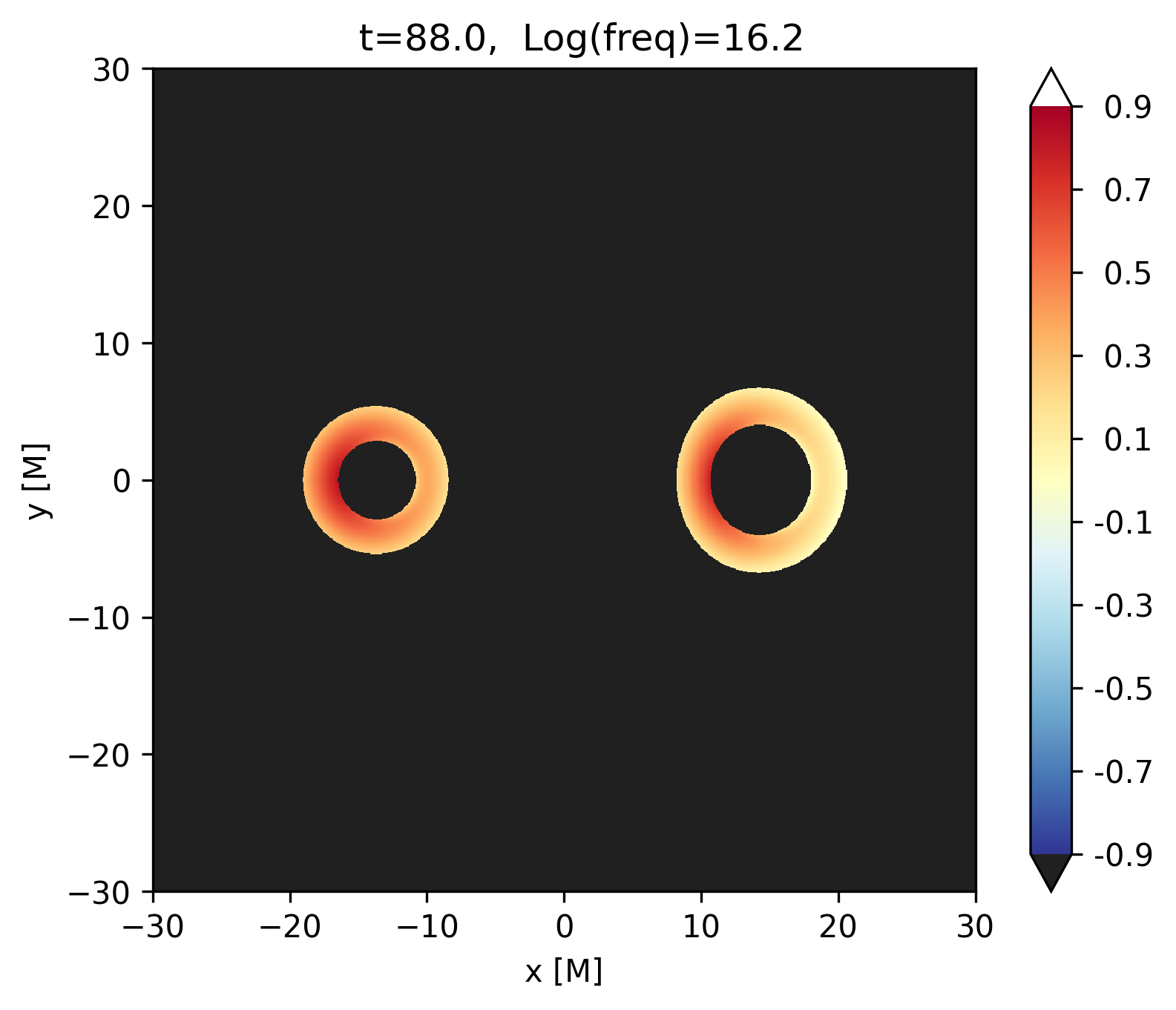}{0.40\textwidth}{(a)}\fig{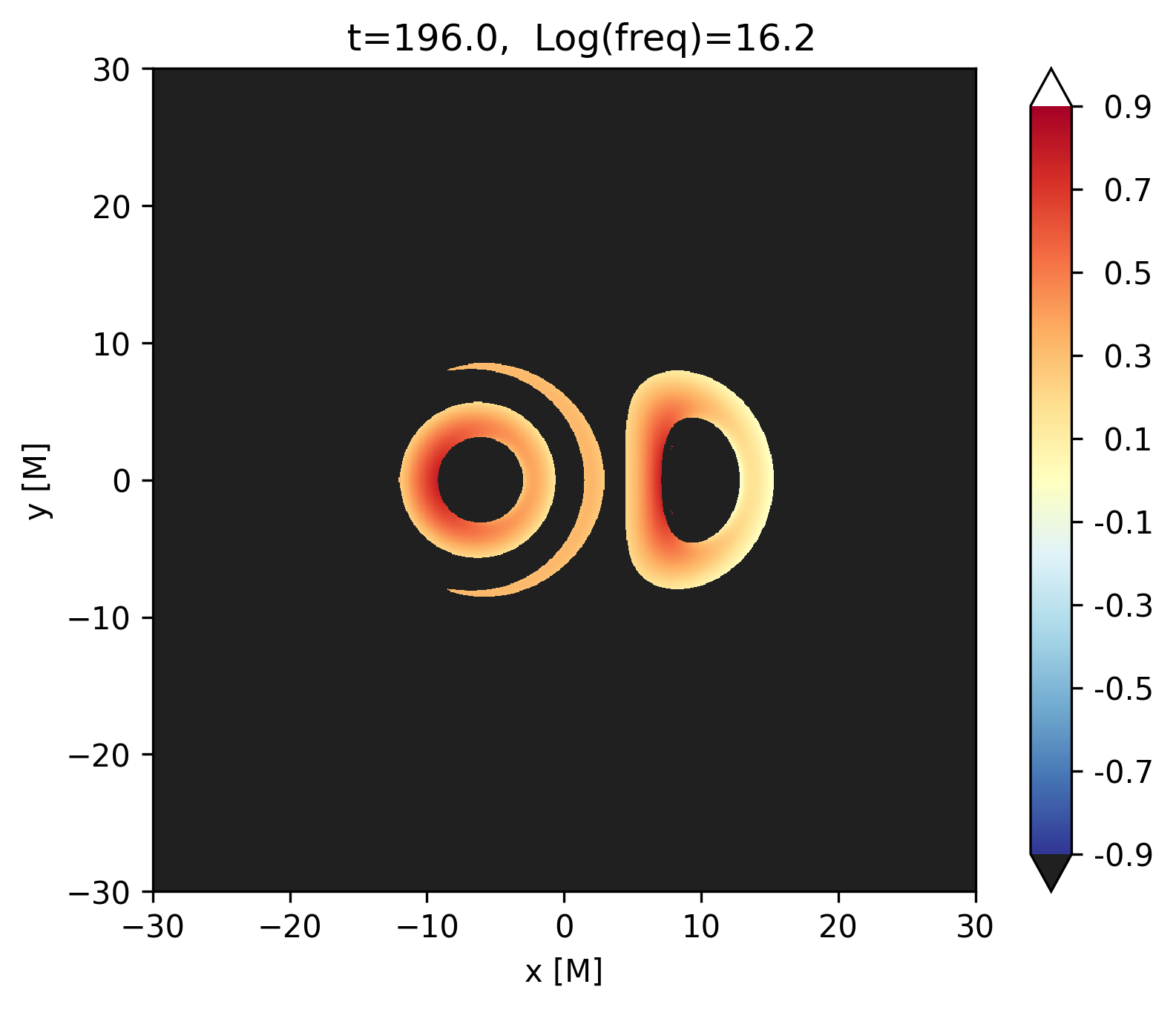}{0.40\textwidth}{(b)}}
\gridline{\fig{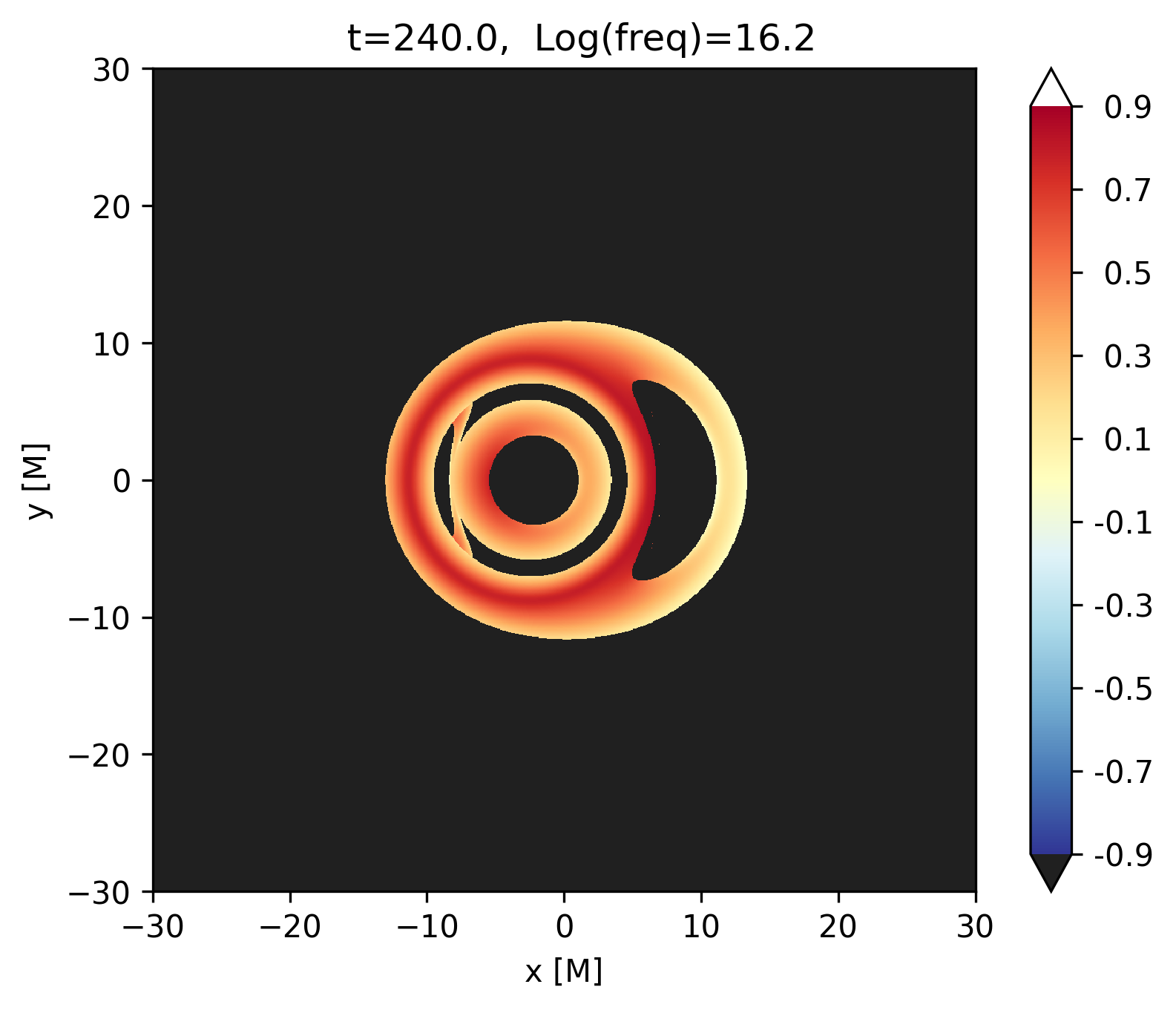}{0.40\textwidth}{(c)}\fig{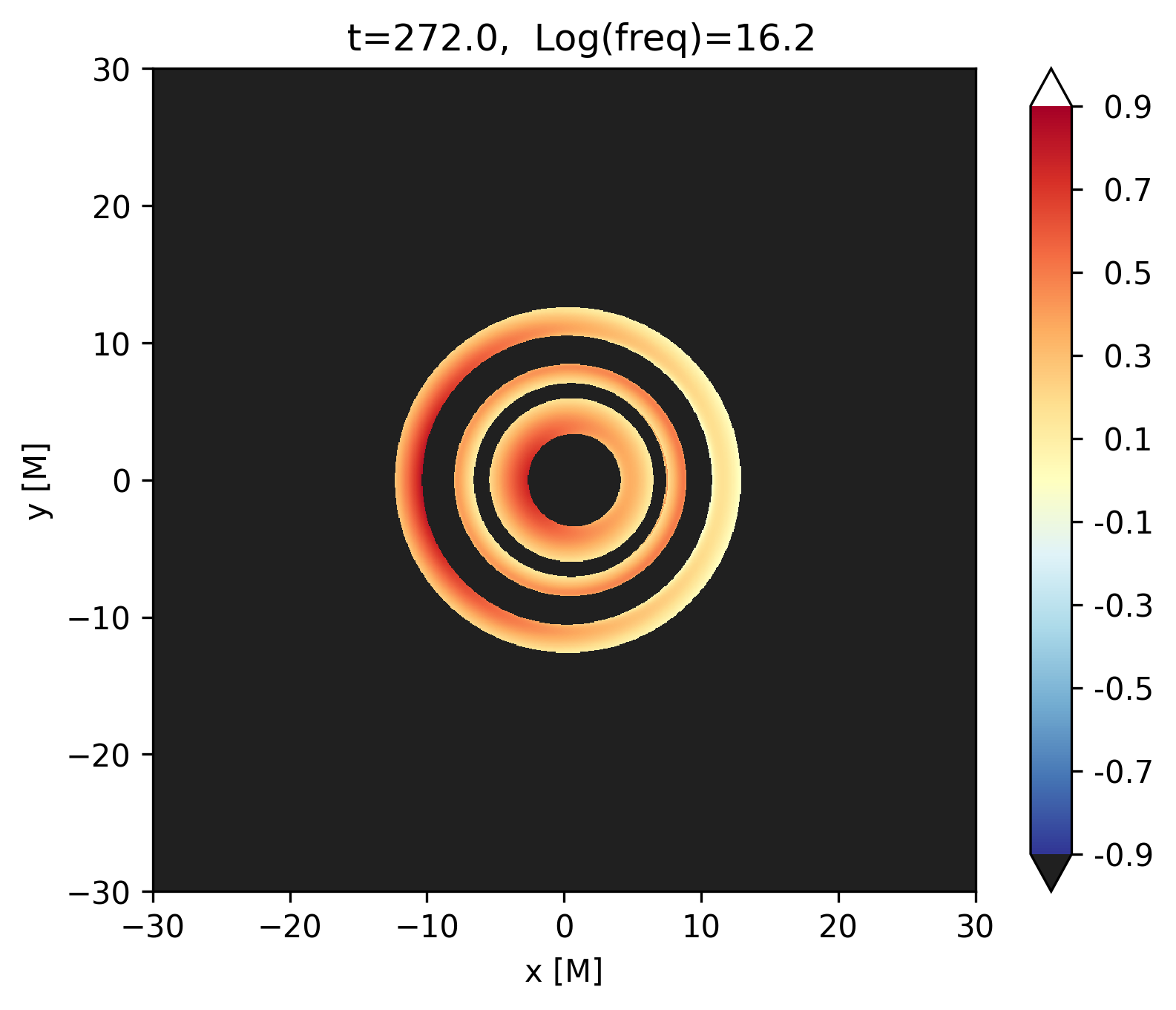}{0.40\textwidth}{(d)}}
\gridline{\fig{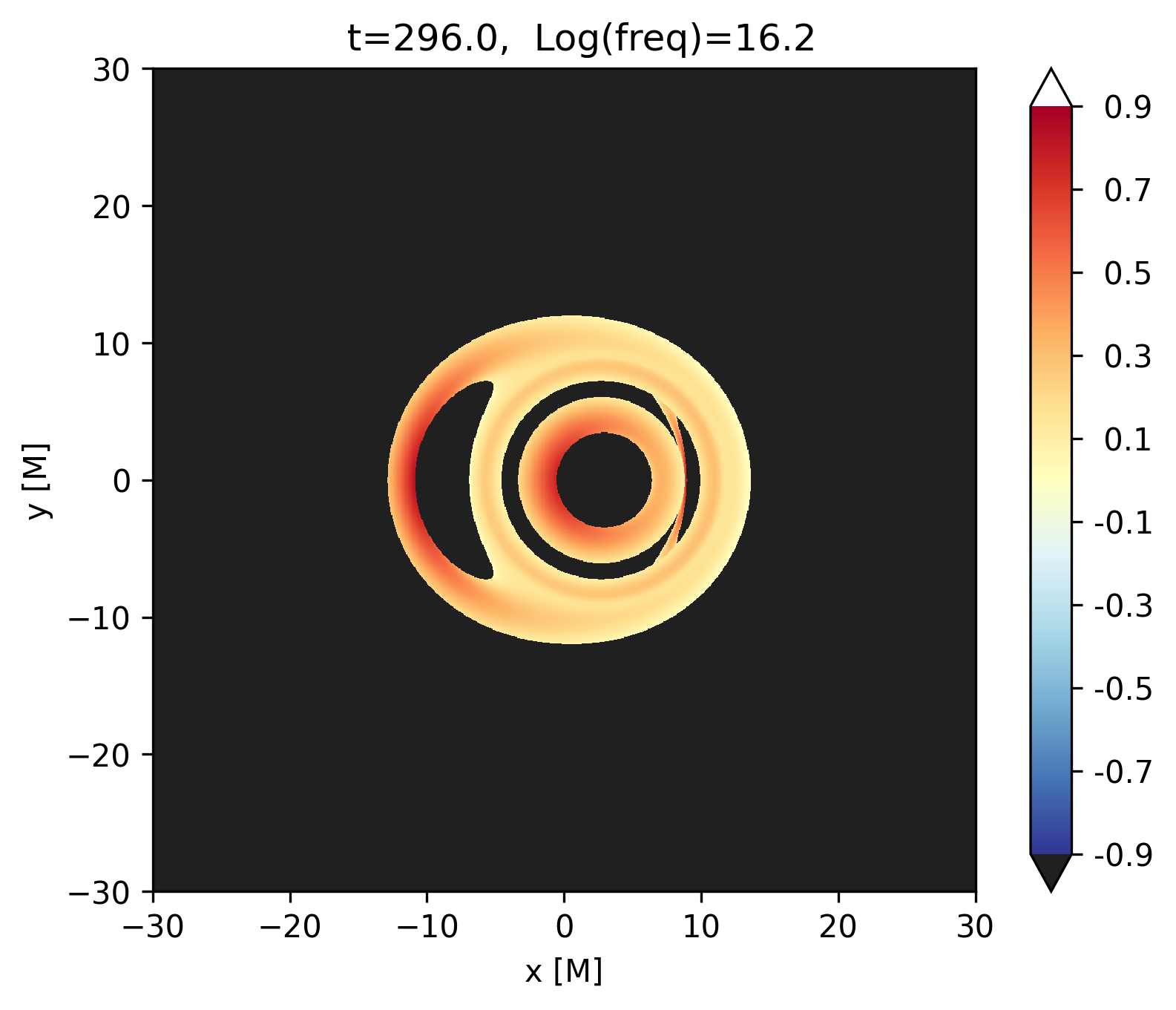}{0.40\textwidth}{(e)}\fig{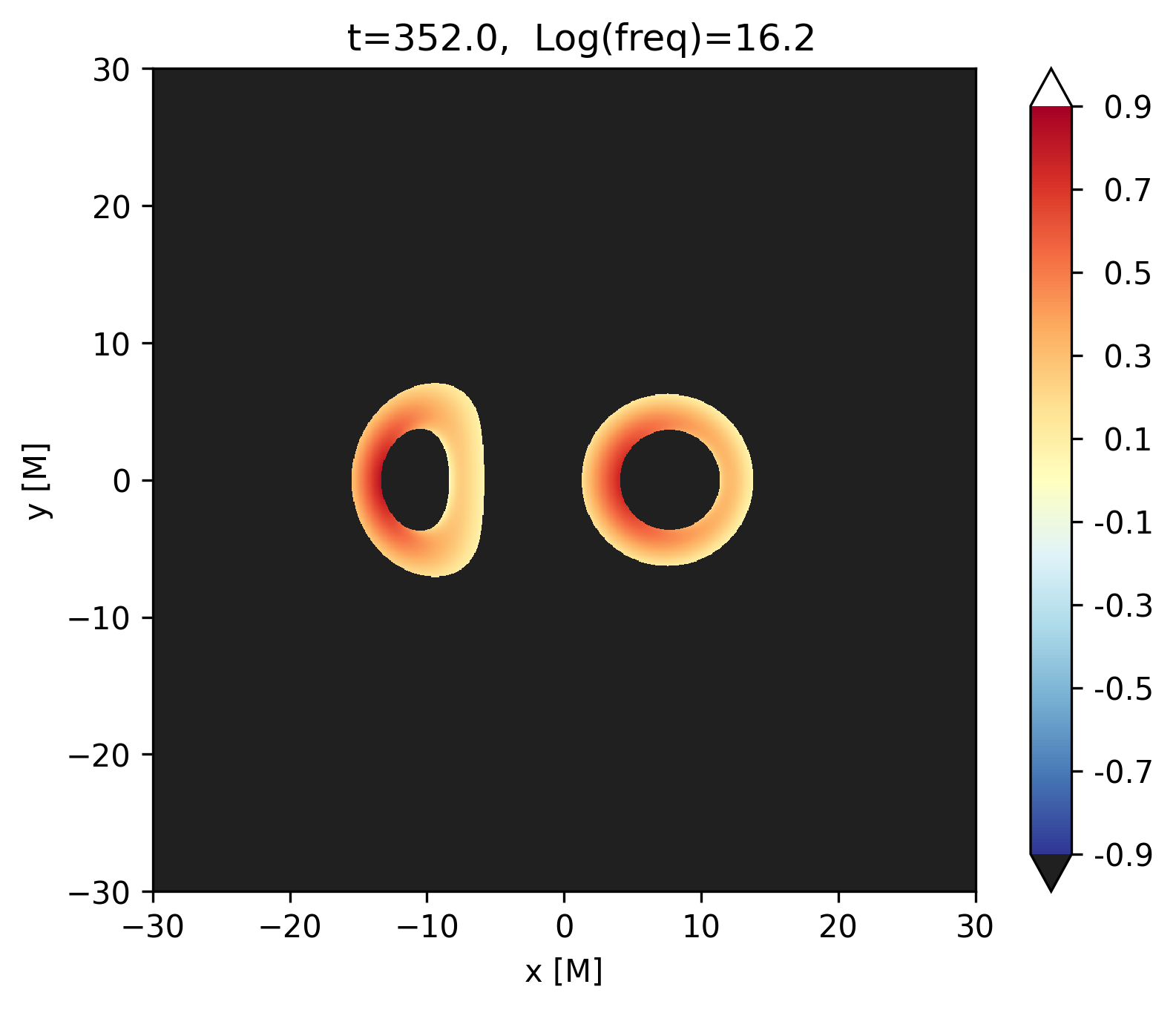}{0.40\textwidth}{(f)}}

    \caption{Panels a - f show images of a binary system at subsequent moments in its orbit. This is the fiducial model (M0) which has the following parameters: $a = 0.0$, $q = 1.0$, $\theta = 90^{o}$, and $\dot{M} = 0.5$. This model was run from $30M$ separation down to $10M$ separation.  The color bar indicates the specific intensity, $I_{\nu}$, in cgs units (erg s$^{-1}$ cm$^{-1}$ Hz$^{-1}$ sr$^{-1}$) on a log scale between $10^{-0.9}$ and $10^{0.9}$ at a frequency of $10^{16}$~Hz. }
    \label{fig:slf}
\end{figure*}

Figure \ref{fig:slf} shows the images and the corresponding observation times in the light curve of the M0 canonical model at six separate times in the binary's orbit. The images show the intensity of light at a frequency of $1.58\times 10^{16}$ Hz. These are edge-on images, $\theta = 90^{\circ}$, of the BHs and their accretion disks. The top half of the rings around the BHs correspond to the light coming from the portion of the disk that is behind the BH with respect to the observer's line of sight. This light is bent due to the strong gravitational environment, so as to appear to be originating from above the BHs. The bottom half of the light rings around the BHs are photons coming from the underside of the accretion disk.

\begin{figure}[h!]
\centering\includegraphics[width=1.0\linewidth]{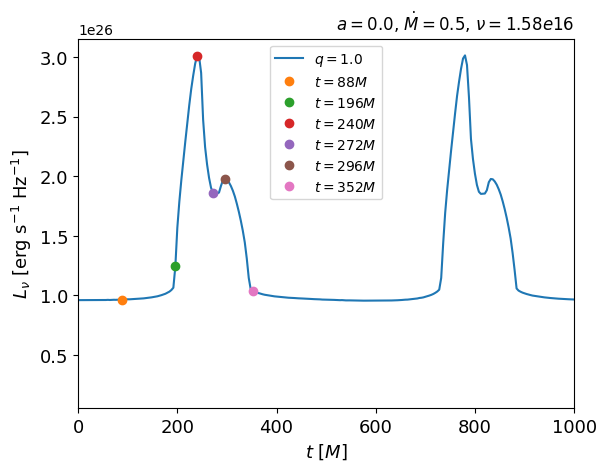}
\caption{Light curve for a single orbit of model M0. The dots show the specific luminosity of the system at the corresponding image times from Fig.~\ref{fig:slf}}\label{fig:slf_lc}
\end{figure}

Figure \ref{fig:slf_lc} is the resulting light curve for M0 which shows two double-peaked flares. These flares are the result of the strong gravitational lensing of light from the binary. When one BH begins to pass in front of the other along the observer's line of sight, it acts as a gravitational lens. These flares occur twice in one orbit of the binary system when the roles of the lens and the source switch between each BH. From Panel (b) of Figure \ref{fig:slf} we can see that the flares occur when one of the BHs begins to lens the other. The gravitational lensing effects cause more light from the accretion disk to be bent toward the observer. The local minimum in the flare occurs when the BHs are directly in opposition to one another and the image produced is a series of nested Einstein rings (Panel (d)). The dip is due to the fact that, when the two black holes are perfectly aligned with the observer (d), the region of maximum magnification is the center of the distant source, i.e., the central gap in the accretion disk. There is an asymmetry in the peaks that is a result of the relativistic beaming effect. The first, more luminous, peak in the flare corresponds to the moment when the lensing BH begins to pass in front of the side of the disk of the source BH that is traveling towards the observer ( Panel (b) \& (c)). The relativistic beaming effect causes light that originates from material traveling toward the observer to appear more luminous. The second smaller peak corresponds to the moment that the lensing BH begins to travel over and lens the side of the disk of the source BH that is traveling away from the observer (Panel (e)).

\subsection{Initial Separation}

\begin{figure}
\centering\includegraphics[width=1.0\linewidth]{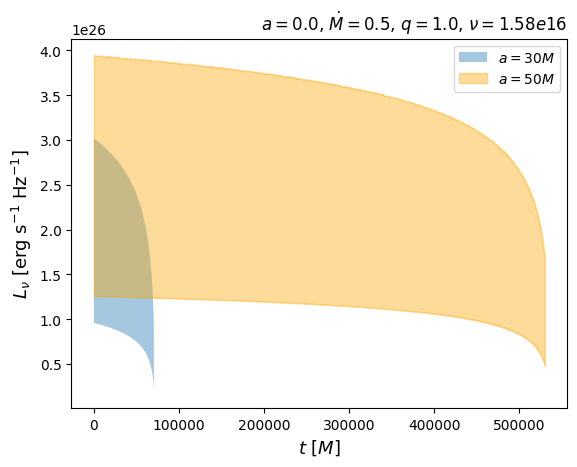}
\caption{Initial separation dependence on the light curves. Shown are the full light curves for model M0 ($r_{12}^{i} = 30M$, blue) and M2 ($r_{12}^{i} = 50M$, orange). }\label{fig:sep_long}
\end{figure}

\begin{figure}
\centering\includegraphics[width=1.0\linewidth]{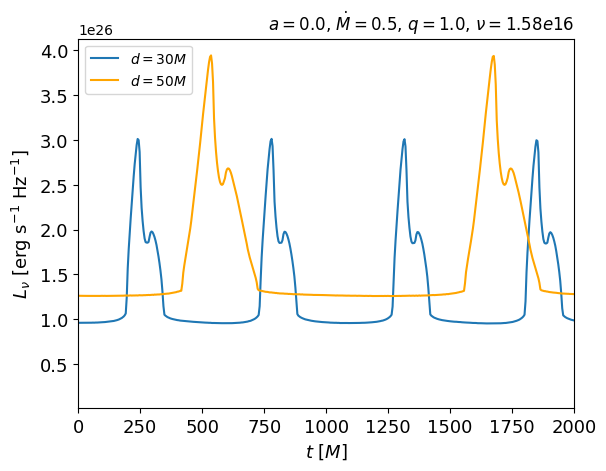}
\caption{Initial separation dependence on the light curves. Shown are the light curves for two orbits of model M0 ( $r_{12}^{i} = 30M$, blue) and for a single orbit of model M2 ($r_{12}^{i} = 50M$, orange). }\label{fig:sep_short}
\end{figure}

Figure \ref{fig:sep_long}  shows the envelope of the total light curve for M0 and M2. Model M0 has an initial separation of 30M and a final separation of 10M. Whereas, Model M2 has an initial separation of 50M and a final separation of 20M. Figure \ref{fig:sep_short} shows the light curves for two models with different initial binary separations for a single orbit of M2 and two orbits of M0. Firstly, the model with an initial separation of $50M$ has a more luminous light curve. The outer radius of each of the mini-disks is proportional to the binary separation. Therefore, the mini-disks extend to larger radii when the BHs are separated by larger distances. The total luminosity of the binary system increases because there is more radiating material. Figure \ref{fig:sep_long} also shows that the light curves for both models decrease in luminosity as the binary inspirals.  As the black holes' separation gets shorter, both of the disks get smaller as well and thus emit less radiation. 

Additionally, Figure~\ref{fig:sep_long} shows that the period of the lensing flares is larger for larger separations. It takes longer for each BH to complete a full orbit. We also note that the flares are wider for $r_{12}=50M$, due to the fact that the disks are larger, and the orbital velocity is slower, thus making the lensing event last longer.

\subsection{Accretion Rate}

\begin{figure}[h!]
\centering\includegraphics[width=1.0\linewidth]{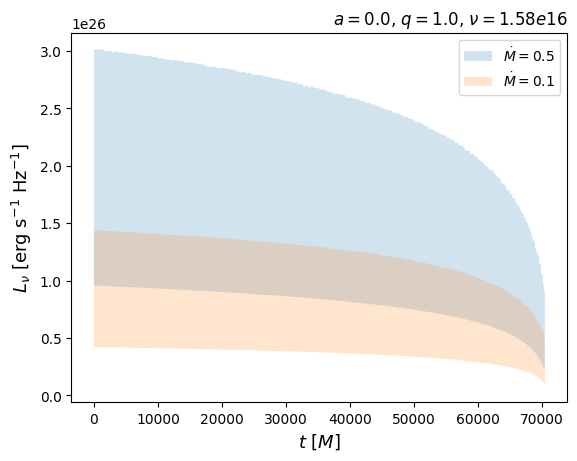}
\caption{Light curves of models M0 (blue), which has an accretion rate of $0.5 \MdotEdd$, and M4 (orange), which has an accretion rate of $0.1 \MdotEdd$. The models start at a binary separation of $30M$ and end at $10M$ separation. Both models have the following parameters: $a = 0.0$, $q = 1.0$, $\theta = 90^{o}$. These light curves show that mini-disks with higher accretion rates are more luminous. }\label{fig:ar_long}
\end{figure}

Model M4 tests the dependence of the light curves on the mass accretion rate of both BHs. In M0, both of the BHs have an accretion rate of $\dot{m} = 0.5 \MdotEdd$, where $\MdotEdd$ is the Eddington accretion rate. In M4, both BHs have an accretion rate of $0.1 \MdotEdd$. Figure \ref{fig:ar_long} shows the envelope of the total light curve for M0 and M4 starting at a binary separation of 30M and ending at a separation of 10M. Not surprisingly, we see that the specific luminosity for the binary system with a larger accretion rate is larger compared to the system with a smaller accretion rate.  A higher accretion rate means that there is more material in the accretion disk that can emit radiation. Therefore, the total light curve for the model with the accretion rate of $0.5 \MdotEdd$ is overall more luminous than the model with the accretion rate of $0.1 \MdotEdd$. 

\begin{figure}[h!]
\centering\includegraphics[width=1.0\linewidth]{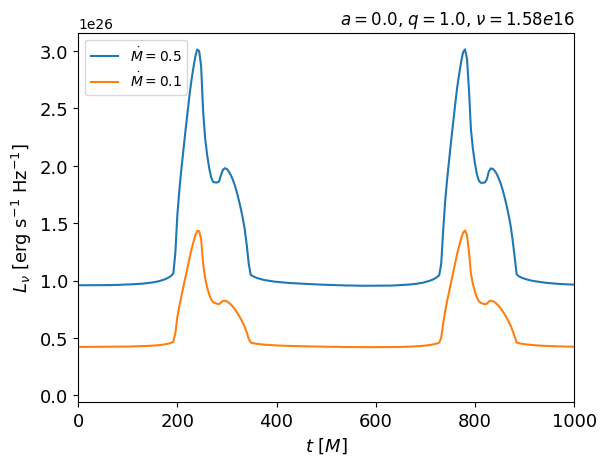}
\caption{Light curves for a single orbit of models M0 ($\dot{M} = 0.5$, blue) and M4 ($\dot{M} = 0.1$, orange).}\label{fig:ar_short}
\end{figure}

Figure \ref{fig:ar_short} shows the light curves for both models between $t = 0M$ and $t = 1000M$, which is about a single orbit of the binary system. Both peaks in the flares for the model with the smaller accretion rate are less prominent compared to the model with the larger accretion rate. Because there is less material producing radiation in the accretion disk, there is less light to be lensed and bent toward the observer. The effect of relativistic beaming, which causes the second peak to be smaller is still present.

We find that the observed flux scales like $F_{\nu} \approx \dot{M}^{1/2}$. Since we assume the thin disk behaves like a blackbody, we can refer to Planck's law at different frequency regimes. In the low-frequency limit of a blackbody spectrum, $F_{\nu} \propto \nu^{2} T$. The temperature of the disk and the mass accretion rate are related by $T \propto \dot{M}^{1/4}$ and therefore $F_{\nu} \propto \dot{M}^{1/4}$ \citep{Novikov:1973}. Near the peak of the spectrum, $h\nu = kT$, which results in $F_{\nu} \propto T^{3}$. In this case $F_{\nu} \propto \dot{M}^{3/4}$. If the observer frequency band falls somewhere between these two regimes, then it is reasonable that $F_{\nu}$ would scale like $\dot{M}^{1/2}$.

\subsection{Mass-ratio}

\begin{figure}[h!]
\centering\includegraphics[width=1.0\linewidth]{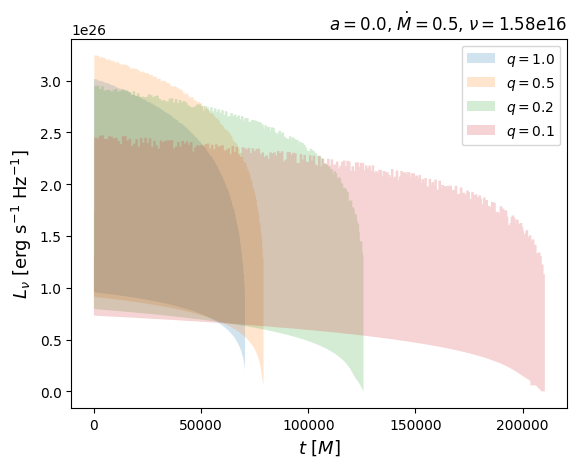}
\caption{Light curves for models M0 ($q = 1.0$, blue), M5 ( $q = 0.5$, orange), M7($q = 0.2$, green), and M9 ($q = 0.1$, red). All models have the following parameters: $a = 0.0$, $\dot{M} = 0.5 \MdotEdd$,$\theta = 90^{o}$. These light curves show that small mass ratio binary systems take longer to inspiral.}\label{fig:mr_long}
\end{figure}

Figure \ref{fig:mr_long} shows the light curves for the four different mass ratios models, $q \in \{1.0 (M0), 0.5 (M5), 0.2 (M7), 0.1 (M9)\}$, for a binary separation of $30M$ down to $10M$ separation and a fixed total mass of $10^{6} \Msun$. The systems with smaller mass ratios take more time to reach the $10M$ separation.  The rate at which a binary system loses energy is $ dE/dt \propto\mu M^3/a^5$ where $\mu = m_1 m_2/M$ is the reduced mass. As the mass ratio decreases, so does the reduced mass. This leads to a slower rate of energy loss and longer inspiral times for smaller mass ratio systems. 

\begin{figure}[h!]
\centering\includegraphics[width=1.0\linewidth]{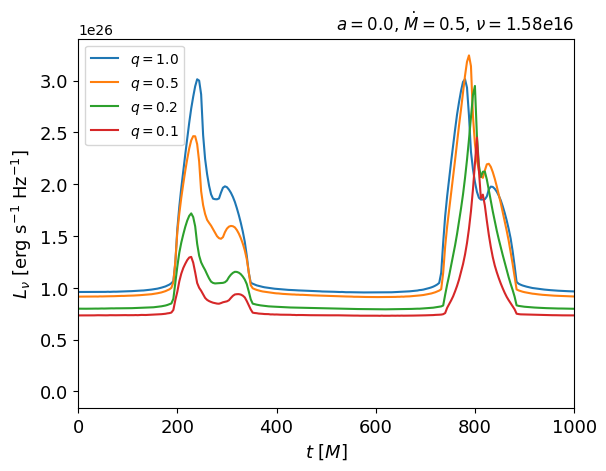}
\caption{Light curves for a single orbit of models M0 ($q = 1.0$, blue), M5 ($q = 0.5$, orange), M7 ($q = 0.2$, green), and M9 ($q = 0.1$, red). The equal mass ratio model (M0) has level flares, whereas the unequal mass models have uneven flares.}\label{fig:mr_short}
\end{figure}

The luminosity of the accretion disks decreases as the mass ratio decreases as well. This is more clearly seen in Figure \ref{fig:mr_short}, which shows the light curves over a single orbit. There are several interesting features of these flares in the light curves. There is a difference in the height of the two lensing flares for each orbit of the non-equal mass BH models. The second flare has a higher amplitude than the first. The smaller flare corresponds to the moment when the smaller mass BH is the lens and the larger mass BH is the source. This is because the gravitational lensing effect is weaker for less massive objects, thus less light is being directed toward the observer.

\begin{figure*}
\gridline{\fig{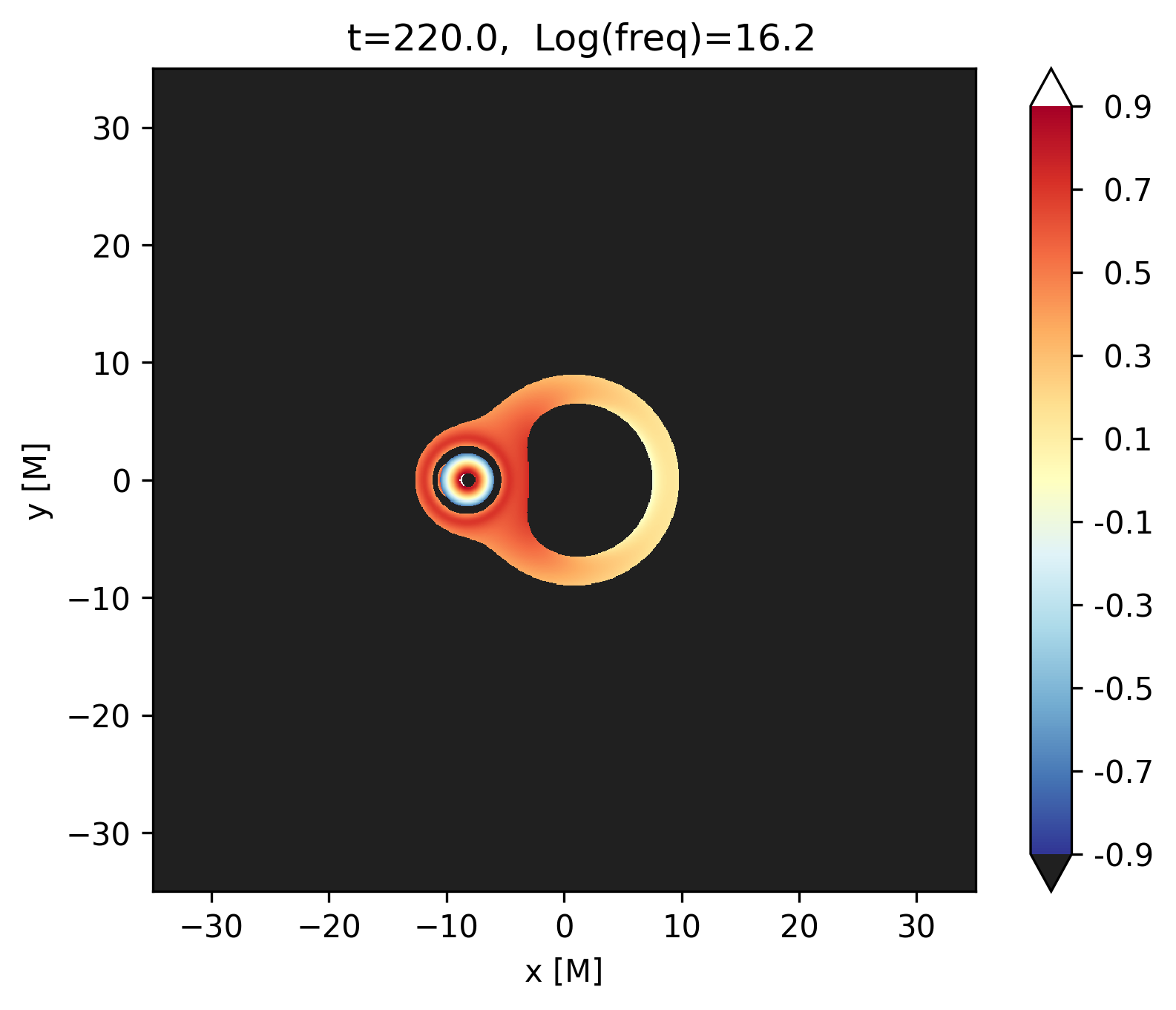}{0.48\textwidth}{(a)}\fig{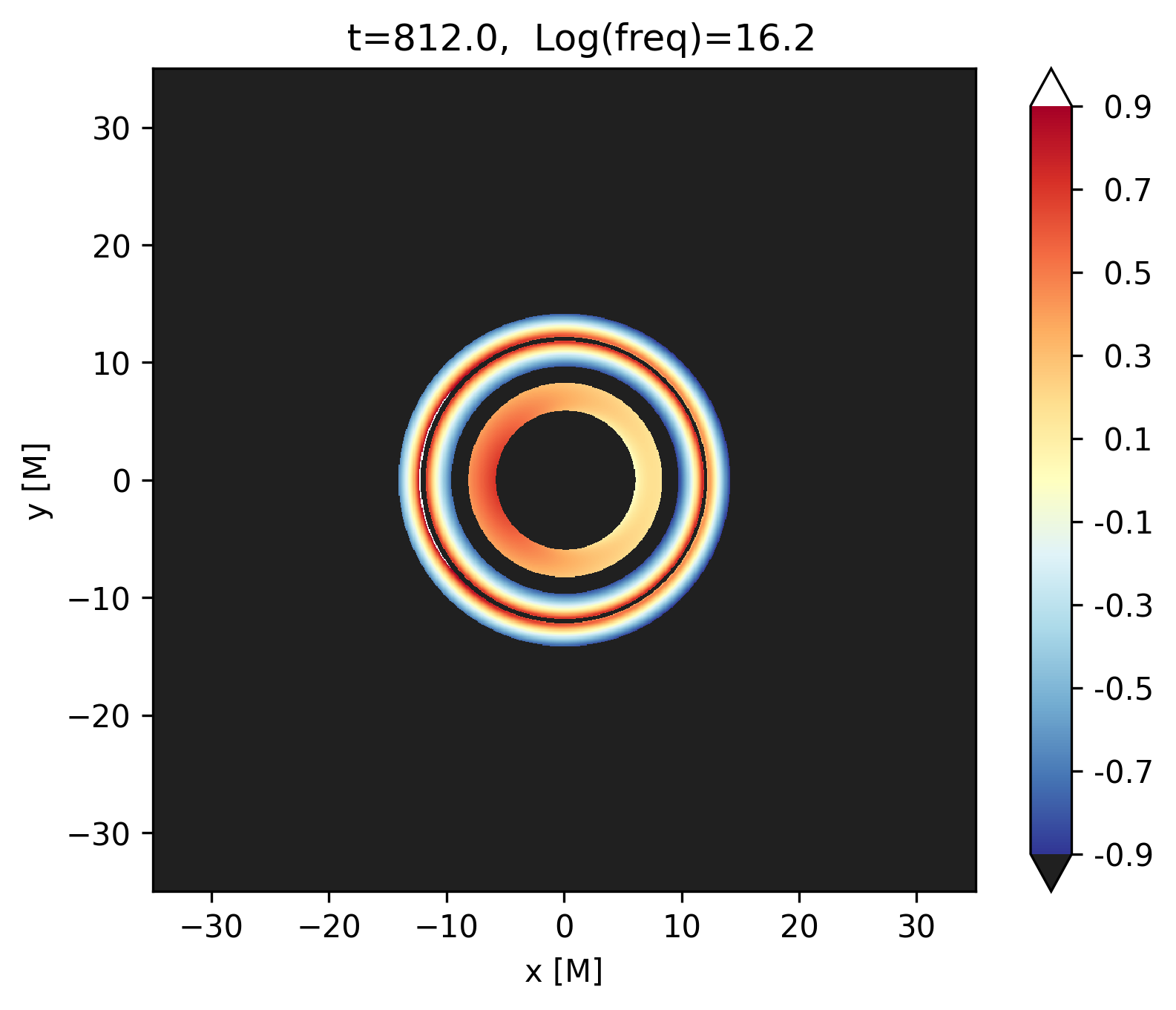}{0.48\textwidth}{(b)}}
\gridline{\fig{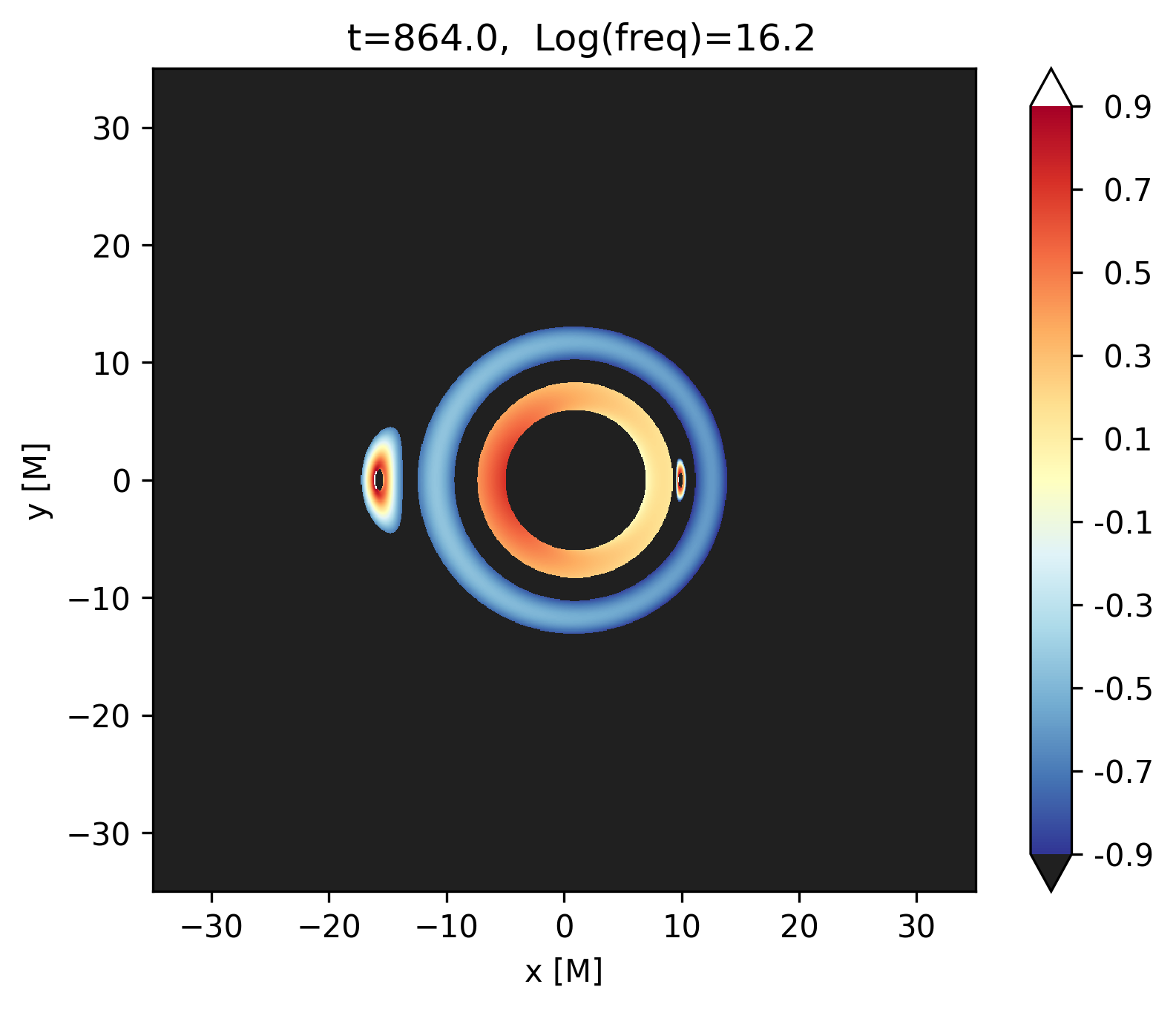}{0.48\textwidth}{(c)}\fig{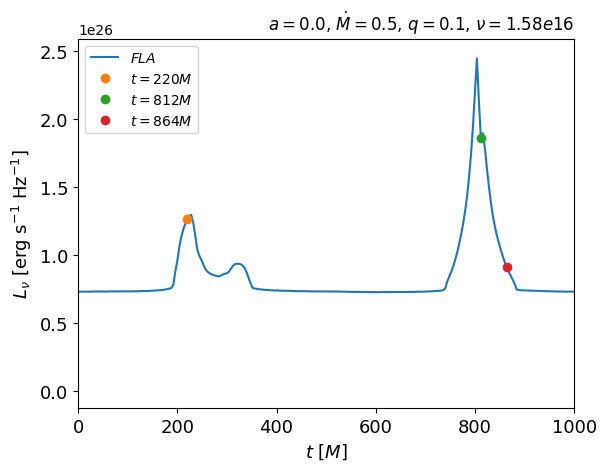}{0.48\textwidth}{(d)}}

    \caption{Images from model M9 ($q = 0.1$) at times $t = 220$ (a), $t = 812$ (b), and $ t = 864$ (c). The light curve with points corresponding to the image times is in panel (d).  In panel (a) the smaller mass BH is lensing the larger mass BH, resulting in a wider but smaller lensing flare. In panel (b) the larger mass BH is gravitationally lensing the smaller mass BH, which causes a more narrow and more intense flare in the light curve. A second image of the source BH can be seen on the right of the lensing BH in panel (c).}
    \label{fig:lc09}
\end{figure*}

Another feature that stands out in the light curve is the fact that the second, larger flare is narrower compared to the first. Figure \ref{fig:lc09} shows the images and light curves of M9 when the larger BH is passing in front of the first. The less massive BH has a smaller accretion disk compared to the more massive BH. This means that the source has a smaller angular size compared to the lens. The lensing from the more massive BH more effectively bends the light from the source so that it converges toward the observer. The greater relative difference in size and lensing strength also results in us being able to see more easily the source BH's second image, which can be seen in Panel c of Figure \ref{fig:lc09} to the right of the lensing BH's image.

Lastly, the second flare for Model M5 ($q = 0.5$) is larger than the second flare for Model M0 ($q = 1.0$). Additionally, the second flare for model M7 ($q = 0.2$) is almost equal in height to model M0. This would not seem at first to physically make sense, since a more massive binary system should be more luminous. However, the second flare in Figure \ref{fig:mr_short} corresponds to the moment when the smaller BH, in the unequal mass models, acts as the source and the larger BH acts as the lens. Moving forward, the flare corresponding to this configuration is referred to as the ``primary flare." Since the smaller BH has a smaller angular size, a larger fraction of its light can be bent and focused by the gravitational lensing effect of the larger BH. This results in an amplification of the observed brightness of the primary flare. This can be explained through the gravitational lensing magnification factor, which depends on the alignment of the source, lens, and observer, as well as the angular size of the source.

For a point source, the Einstein radius, which describes the scale of the gravitational lensing effect, is given by

\begin{align}
    \theta_{E} = \sqrt{\frac{4GM}{c^2}\frac{D_{LS}}{D_{L}D_{S}}} \ ,
\end{align}
where $D_{S}$ is the distance from the observer to the source, $D_{L}$ is the distance from the observer to the lens, $D_{LS}$ is the distance from the lens to the source, and $M$ is the mass of the lensing object. This quantifies the angular deflection of the light. The magnification factor of the light from the source is given by

\begin{align}
    \mu = \frac{\theta_{E}}{\beta} \frac{d\theta_{E}}{d\beta} \ ,
\end{align}
where $\beta$ is the angular separation of the source from the lens's center. The magnification factor depends on the angular size of the source in relation to the Einstein angular radius. The magnification factor increases as the angular size of the source decreases, leading to an amplification of the primary flare for smaller mass ratios.

\subsection{Spins}

\begin{figure}
\centering\includegraphics[width=1.0\linewidth]{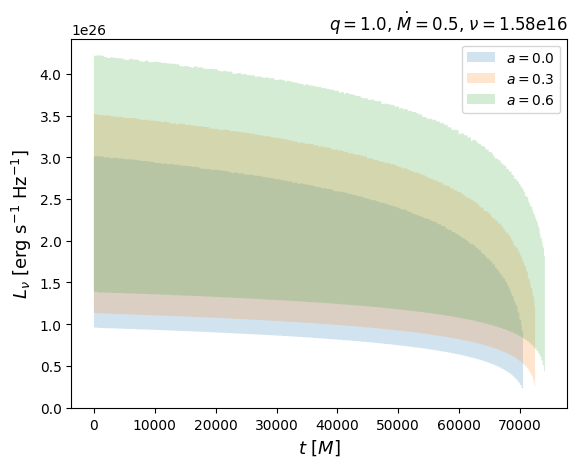}
\caption{Light curves from 30M separations down to $10M$ separation for models M0 ($a = 0.0$, blue), M11 ($a = 0.3$, orange), and M13 ($a = 0.6$, green). 
The higher the spin of both BHs, the more luminous the light curve.}\label{fig:spin_long}
\end{figure}

\begin{figure}
\centering\includegraphics[width=1.0\linewidth]{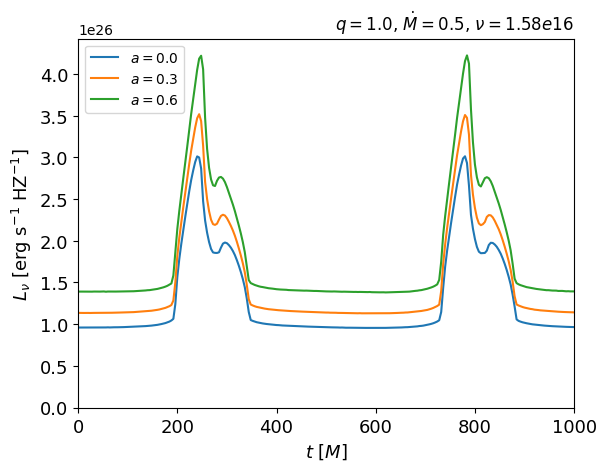}
\caption{Light curves for a single orbit of models M0 ($a = 
 0.0$, blue), M11 ($a = 0.3$, orange), and M13 ($a = 0.6$, green) showing the effects of the spins of both BHs. }\label{fig:spin_short}
\end{figure}

Models M11 and M13 test the spin dependence of the light curves. From both Fig.~\ref{fig:spin_long} and Fig.~\ref{fig:spin_short} we can see that the overall luminosity increases as the spin parameter increases. This is because, as the spin of the black hole increases, the radius of the innermost stable circular orbit decreases. In addition, the amount of energy per unit mass liberated through accretion increases with increasing spin. Both spin models have significantly higher peak luminosity compared to the other models tested in this paper. The peak (specific) luminosity for model M11 ($a =0.3)$ is around $3.5 \times 10^{26} ~ {\rm erg~s}^{-1}{\rm Hz}^{-1}$, and for model M13 ($a = 0.6$) is above $4 \times 10^{26} ~ {\rm erg~s}^{-1}{\rm Hz}^{-1}$. Additionally, we see the two sub-peaks in each flare move slightly closer together with increasing spin, as the inner radius of the disk decreases, pushing the approaching and receding edges closer together.

\subsection{Viewing Angle}

\begin{figure}
\centering\includegraphics[width=1.0\linewidth]{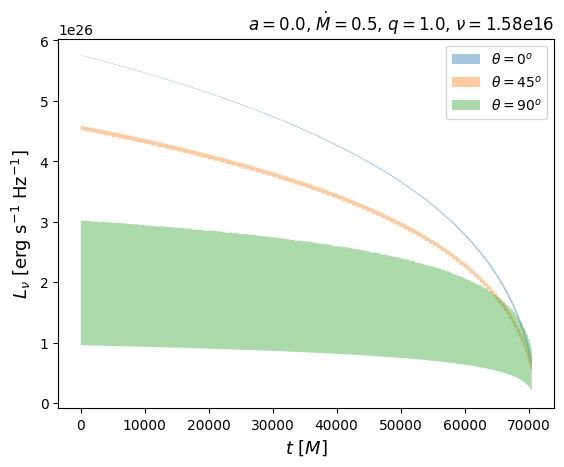}
\caption{Lightcurves for three different viewing angles for binaries starting at 30M separation and inspiraling down to 10M separation. The light curve for model M0 ($\theta = 90^{\circ}$) is in green, model M15 ($\theta = 0^{\circ}$) is in blue, and model M16 ($\theta = 45^{\circ}$) is in orange.}\label{fig:va_long}
\end{figure}

\begin{figure}
\centering\includegraphics[width=1.0\linewidth]{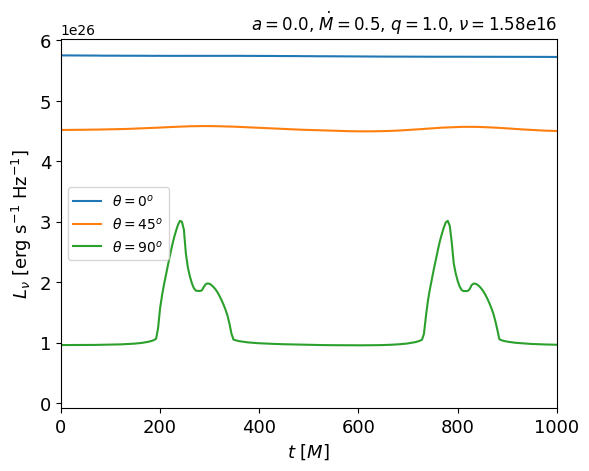}
\caption{Light curves for a single orbit of models M0 ( $\theta = 90^{\circ}$, green), M15 ($ \theta = 0^{\circ}$, blue), and M16 ($\theta = 45^{\circ}$, orange).}\label{fig:va_short}
\end{figure}

Models M15 and M16 test the viewing angle dependence. Figure~\ref{fig:va_long} and Figure~\ref{fig:va_short} show the full and single orbit light curves for these models respectively. We see that the lensing effects that show up on the light curves diminish as the viewing angle decreases. With a viewing angle of $0^{\circ}$, the BHs are not passing in front of each other along the observer's line of sight. There is no gravitational lensing taking place in this model and the lack of lensing flares in the light curves reflects that, while the overall flux increases with decreasing inclination, simply due to geometric effects: the observer can see a larger solid angle of the disk when viewed face-on.

\subsection{The Fast-light Approximation}

\begin{figure}[h!]
\centering\includegraphics[width=1.0\linewidth]{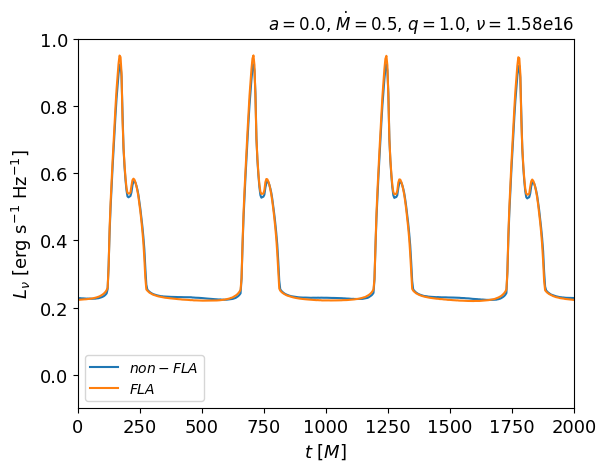}
\caption{Light curves for two orbits of binary systems with $q = 1.0$ for FLA (M0) and non-FLA (M1) models. The light curves were aligned by calculating the max overlap.}\label{fig:M0vM1}
\end{figure}

Figure~\ref{fig:M0vM1} shows the light curves for a single orbit for models M0 and M1. M1 has the same parameters as M0; however, the fast-light approximation is dropped for this model. The light curves are very similar in shape, but the FLA model has flares that have a slightly higher amplitude compared to the non-FLA model. This is because when the BHs are allowed to move during the geodesic integration, i.e. when the FLA is dropped, not all of the photons that leave the disk at the same time also reach the observer at the same time, which causes a decrease in the luminosity of the flares. 

\begin{figure*}
\gridline{\fig{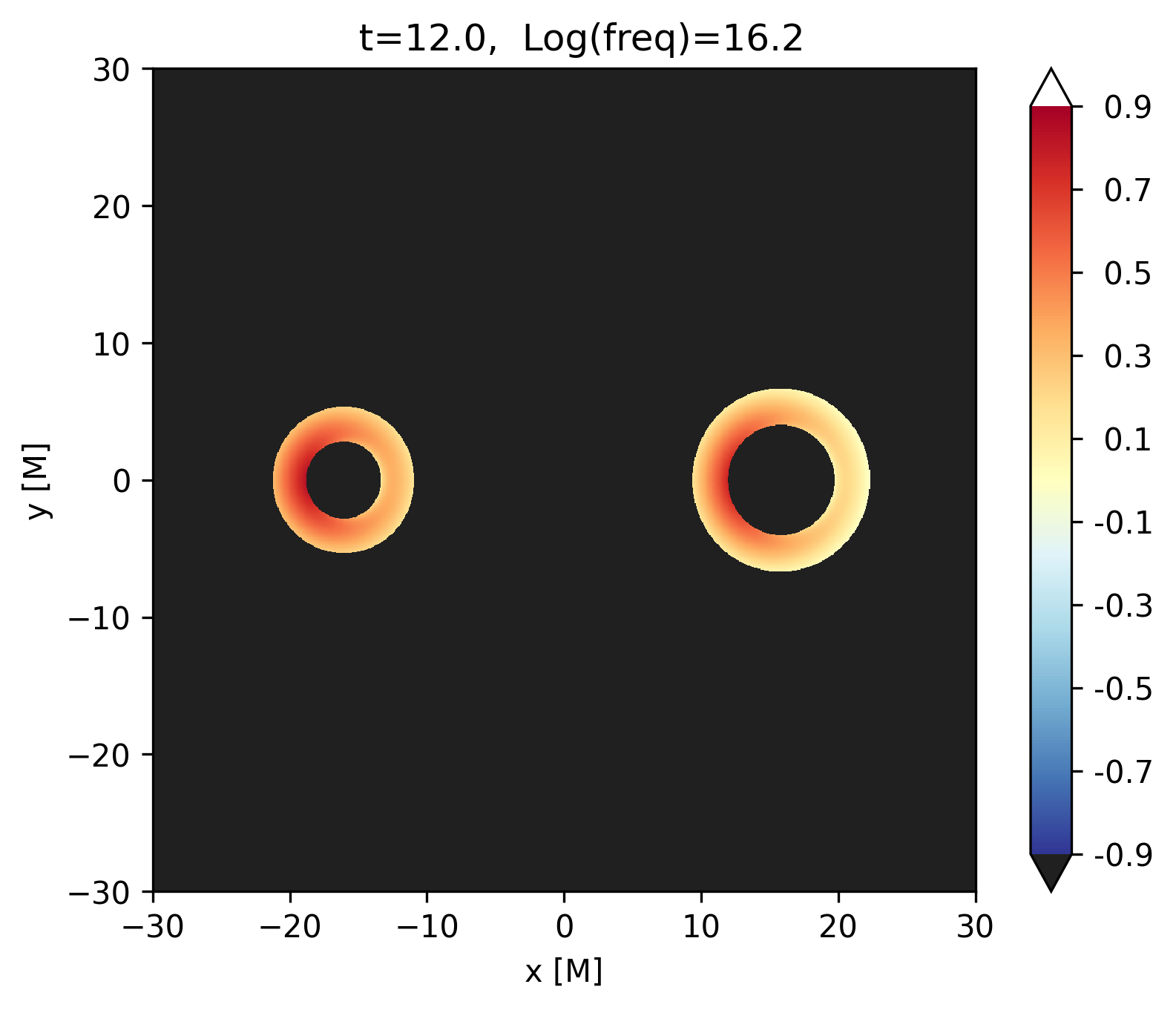}{0.48\textwidth}{(a) $q = 1.0$, FLA }\fig{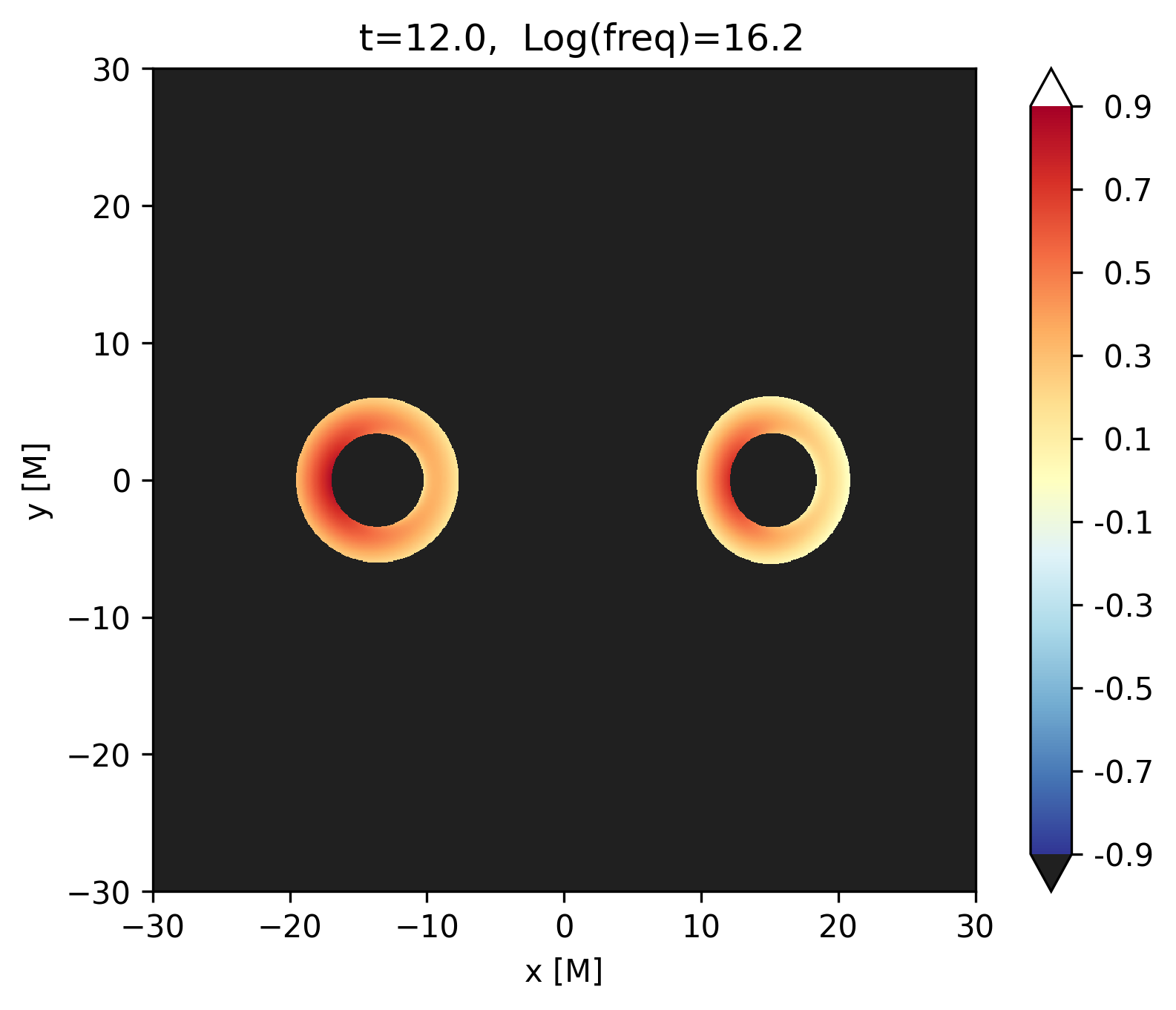}{0.48\textwidth}{(b) $q =1.0$, non-FLA}}
\gridline{\fig{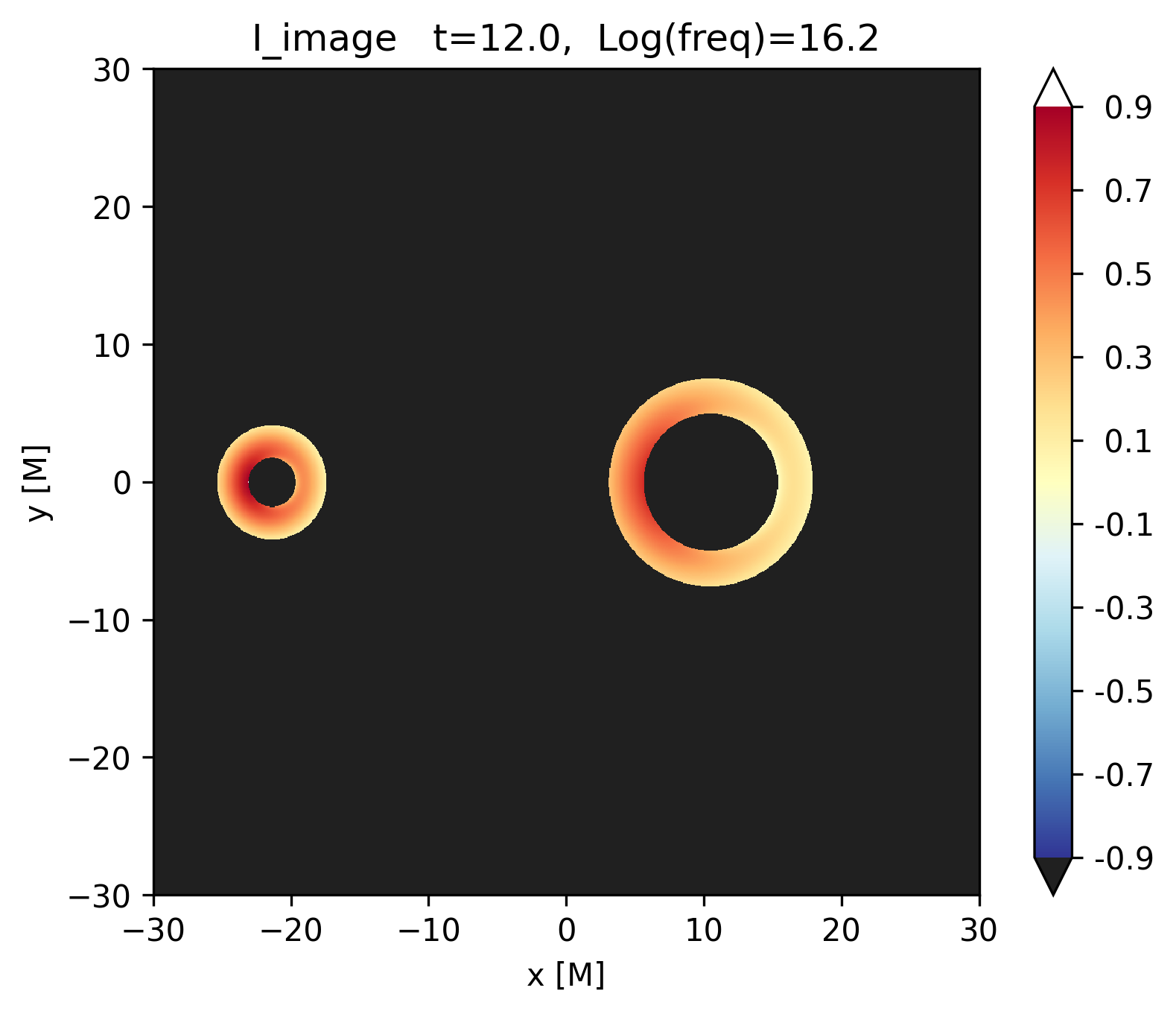}{0.48\textwidth}{(c) $q = 0.5$, FLA }\fig{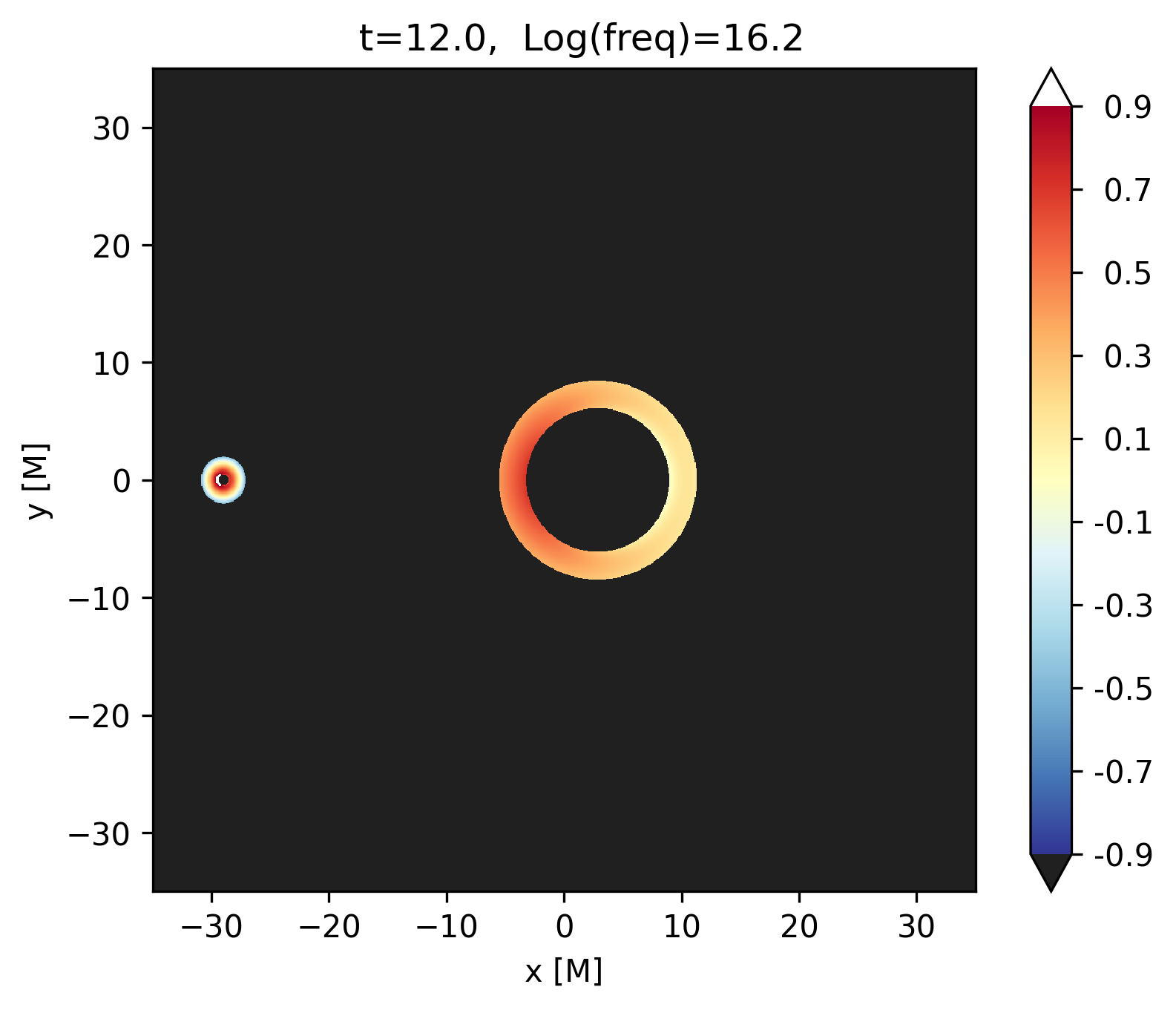}{0.48\textwidth}{(d) $q = 0.1$, FLA}}
    \caption{Images of the binary system at $t = 12.0M$ for four different models. Each image has a color bar on the right for the specific intensity of the accretion disks. The color bar is a log scale from $10^{-0.9}$ to $10^{0.9}$ at a frequency of  $\approx 10^{16}$ Hz. Model M0 (a) is an equal mass binary system and uses the fast light approximation. Model M1 (b) is an equal mass binary system and does not use the fast-light approximation. Model M5 (c) has a mass ratio of 0.5 and uses the fast-light approximation. Model M9 (d) has a mass ratio of 0.1 and uses the fast light approximation. The image of the equal mass binary system with the fast-light approximation shows BHs that are not equal in size. The image of the equal mass binary system that does not use the fast-light approximation shows BHs that are equal in size. Additionally, unequal mass binary systems show a size difference in the images of the BHs. It is possible by looking at just the modeled images of the binary system to confuse an equal mass binary system under the fast-light approximation with an unequal mass binary system. However, the light curves for an equal mass binary are easily distinguishable from an un-equal mass binary (Figure \ref{fig:mr_short}).}
    \label{fig:EventHorizonComp}
\end{figure*}

One major question is whether or not the use of the fast-light approximation could cause features in the images and light curves that could mislead the inference of physical parameters. For an equal mass binary system, the fast light approximation causes the approaching mini-disk to appear smaller compared to the mini-disk that is receding away from the observer. Panel a of Figure~\ref{fig:EventHorizonComp} shows an image of the mini-disks for model M0 at a camera time of 12M. There is a clear difference in size between the mini-disks. Panel~b shows the BHs at the same camera time for model M1, which drops the FLA, and the mini-disks appear more comparable in size. Panels c and d are of models M5 and M9 respectively, which are models with smaller mass ratios. The approaching mini-disks in these images appear smaller than the receding mini-disks because they actually are smaller in size. Therefore, a simulation that uses the FLA could produce images of a binary system that could be confused with one with a mass ratio smaller than 1. However, this issue can be resolved by looking at the light curves that these systems would produce (Figure \ref{fig:mr_short}). The light curve for the FLA and equal mass model has flares of equal height. The flares have the same amplitude regardless of which BH is the lens and which is the source. This is not the case for any of the models with mass ratios less than 1. As the mass ratio decreases, the difference in flare amplitude becomes more pronounced. Therefore, although the images of simulations using FLA appear similar to small mass-ratio simulations, their light curves do not. 

\begin{figure}[h!]
\centering\includegraphics[width=1.0\linewidth]{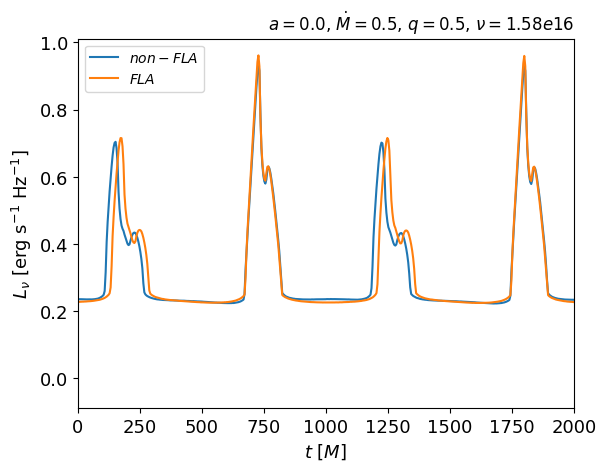}
\caption{Light curves for two orbits of binary systems with $q = 0.5$ for FLA (M5) and non-FLA (M6) models. The light curves were aligned by calculating the max overlap. This method resulted in the primary flares aligning between models and a delay in the arrival of the secondary flares for the FLA model.}\label{fig:M5vM6}
\end{figure}

\begin{figure}[h!]
\centering\includegraphics[width=1.0\linewidth]{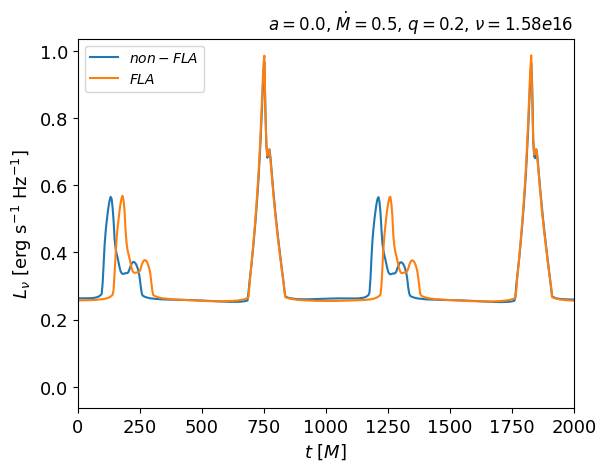}
\caption{Light curves for two orbits of binary systems with $q = 0.2$ for FLA (M7) and non-FLA (M8) models. The light curves were aligned by calculating the max overlap. This method resulted in the primary flares aligning between models and a delay in the arrival of the secondary flares for the FLA model.}\label{fig:M7vM8}
\end{figure}

\begin{figure}[h!]
\centering\includegraphics[width=1.0\linewidth]{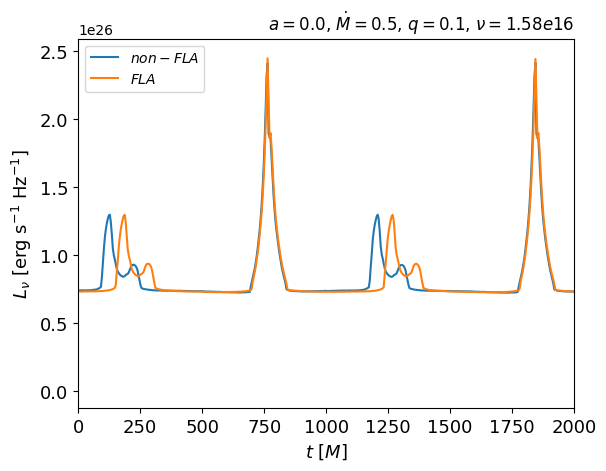}
\caption{Light curves for two orbits of binary systems with $q = 0.1$ for FLA (M9) and non-FLA (M10) models. The light curves were aligned by calculating the max overlap. This method resulted in the primary flares aligning between models and a delay in the arrival of the secondary flares for the FLA model.}\label{fig:M9vM10}
\end{figure}

Figures \ref{fig:M0vM1}, \ref{fig:M5vM6}, \ref{fig:M7vM8}, and \ref{fig:M9vM10}  compare the light curves between the FLA and non-FLA models for the four different mass ratios that we tested. In order to ensure that we are comparing the FLA and non-FLA models at the same time in their orbit, the non-FLA light curves have to be advanced in time by approximately $1000M$. This is necessary because, for non-FLA models, the light seen at the camera has left the mini-disks earlier by the amount of time it takes the light to travel to the camera, which is about $1000M$ in time because the camera lies at $r_\mathrm{cam}=1000M$ from the center of mass (COM).

A max overlap analysis was then performed over the time interval of a single orbit between the FLA and non-FLA light curves for each mass ratio. This calculation resulted in the alignment of all of the lensing flares for the equal-mass binary system and the alignment of the primary lensing flares in the unequal-mass systems. With the primary flares aligned for the entirety of the inspiral for each unequal mass model, we see that there is a time difference in the arrival time of the secondary flares between the FLA and non-FLA models. The non-FLA model secondary flares occur before the FLA flares. This effect is due to the finite speed of light and the relative positions and velocities of the black holes with respect to each other. For small $q$, the primary black hole's orbital radius and thus orbit is smaller compared to that of the secondary black hole. During the time when the black holes are approaching alignment with respect to the observer, the primary black hole does not have to travel very much in its orbit, while the secondary black hole must cover a significant portion of its orbit in order to align with the observer and produce the lensing flare (see Appendix). We have seen that the primary peak corresponds to when the larger black hole is acting as the lens and the secondary flare corresponds to the secondary black hole acting as the lens. For the non-FLA run, the orbital configuration of the BHs is relatively the same as the configuration for the FLA run, since the lens's position does not change significantly during the light's travel time between the source and the lens. However, for the secondary flare, a time delay must be considered because the light from the larger black hole encounters the faster-moving secondary black hole, which travels a greater distance during the light travel time between the two black holes. Thus, in the non-FLA run, the secondary flare corresponds to light emitted from the larger black hole at an earlier orbital configuration than in the FLA. This results in a shorter time interval between the primary and secondary flares in the non-FLA, explaining the time difference in the light curves. This is further demonstrated in the Appendix.

\section{Discussion \& Conclusions}\label{sec:discussion}

This paper presents a new analytical accretion disk model for the mini-disks surrounding each BH in an SMBHB system. The model improves upon the Novikov--Thorne thin disk model by including radiation from within the ISCO. GRMHD simulations of the gas around supermassive black holes show that the radiation within the ISCO significantly impacts the spectra and light curves of these systems. An analytical disk model greatly reduces the computational time needed to simulate the light curves and images of these binary systems. This time reduction allows for a large parameter space study to be conducted. In this paper, we present the results of the parameter study. 

This work investigates how the accretion rate, the mass ratio, the spins of the BHs, the viewing angle, the initial binary separation, and the fast-light approximation affect the light curves and images produced by our analytical disk model. The following are the main results of this paper:
\begin{itemize}
\item The light curves of binary systems that are viewed edge-on have periodic flares due to the gravitational lensing of the light from the accretion disks. The flares occur twice per orbit of the binary system.
\item Each flare caused by gravitational self-lensing has two peaks. The first peak is always stronger than the second due to the differential relativistic beaming effect between the part of the disk moving towards the observer, producing the first peak, and the part of the disk that is moving away from the observer, producing the second peak. 
\item Unequal mass binary systems have flares of unequal height for each orbit. This asymmetry is due to the fact that less massive BHs are less effective gravitational lenses compared to more massive BHs. 
\item Some unequal mass binary systems can be more luminous than equal mass binary systems while the more massive black hole is acting as the lens. 
\item Higher spin parameters result in light curves with a larger maximum luminosity, due to the location of the innermost stable circular orbit.
\item The use of the fast-light approximation can give the appearance of a mass ratio less than unity in the images of an equal mass binary system, however, the light curves have lensing flares of equal height which are not characteristic of unequal mass binary systems.
\item When comparing light curves between FLA and non-FLA models for unequal mass binary systems, there is a delay in the secondary flare for the FLA models.
\end{itemize}

These results demonstrate that while the fast-light approximation simplifies and expedites the calculation of light curves and images, it still captures the essential features and trends observed in more physical models that do not use this approximation, validating its use in GRMHD simulations and in post-processing light curves and spectra from these simulations. However, it is important to consider that the delay between the flares in the FLA calculations could cause a difference in the temporal power spectrum analysis. The temporal power spectrum of a signal reflects the distribution of power across different frequencies over time, and the precise timing of flares can influence the resulting spectrum. Delays introduced in FLA models could lead to variations in the power spectrum that might not be present in the non-FLA models, potentially affecting the interpretation of the underlying physical processes. We leave the exploration of this effect for future work.

\cite{SLF1,SLF2}  have previously conducted a similar parameter space exploration for an analytical mini-disk model. 
The main difference in our study is that we employ a so-called ``superposed PN" spacetime, i.e. a spacetime with two superposed and boosted black hole metrics in harmonic coordinates, and the orbits of the BHs are found by solving the 3.5PN-order equations of motion, allowing us to accurately consider inspiraling binary orbits down to a 10M separation. In contrast, \cite{SLF1,SLF2} use a superposition of two Cartesian Kerr--Schild metrics, where the BHs follow Keplerian orbits. This approach does not accommodate inspiraling binary systems and restricts the light curve study to one orbit of the binary.  The way we superpose two Kerr-Schild spacetimes is also different.  In addition to the coordinate transformation from Kerr-Schild coordinates to Cook-Schild harmonic coordinates, which is necessary for using the PN trajectories expressed in harmonic coordinates, we also boost the black holes to capture their motion in the inertial reference frame of the system. Although \cite{SLF1,SLF2} also drop the fast light approximation for their study, they do not systematically investigate the differences in the light signals with and without the approximation. We performed a comprehensive comparison of the FLA and non-FLA for all parameters we investigated, demonstrating that while the FLA simplifies calculations and retains essential trends in the light curves, the non-FLA approach would capture the timing and frequency content more accurately. We further see asymmetric flare profiles from the light curves, while they find asymmetric flares for only the FLA cases and symmetric flares for the non-FLA cases.  We identify the cause of the asymmetry in our flares arising from the natural asymmetry of each mini-disk because the side that enters the lensing region first is the approaching, and thus brighter, side of the background mini-disk.  We cannot say why they do not see asymmetric flares for the non-FLA case, but it may be due to differences in how the Doppler shift is calculated.  Because we are in the relativistic regime, we make no approximations on how emission is calculated and always calculate emission in the local fluid frame (deboosted and transformed to local KS coordinates) and transport the radiation using Lorentz-invariant means, as described prior. 

In the future, we plan to extend this parameter space study by calculating the spectra of these models as well. Our analytical disk model can be improved upon by adding a model of the CBD. The CBD would be superimposed onto our current mini-disk model using the NT model already implemented within \Bothros.  Following this addition of the CBD, a larger parameter study will be launched. This analytical model will greatly facilitate the study of the parameter space of supermassive binary black hole systems and will allow us to focus the computationally expensive GRMHD simulations on interesting areas of the parameter space based on these results.

\section*{Acknowledgements}\label{}

K.P., S.C.N., M.C., J.S., and B.J.K. gratefully acknowledge NASA for financial support from NASA Theory and Computational Astrophysics Network (TCAN) Grant No. 80NSSC24K0100 to RIT and GSFC. K.P. and M.C. also acknowledge the National Science Foundation (NSF) for financial support from Grants No. PHY-2110338, No. OAC-2004044/1550436/2004157, No. AST-2009330, No. OAC-1811228, No. OAC-2031744 and No. PHY1912632. 
E.M.G. acknowledges funding from the National Science Foundation under grant No. AST-2108467 and from an Institute for Gravitation and the Cosmos fellowship. J.P. acknowledges support from a CONICET fellowship.
Additionally, K.P. received partial support from GSFC, E.M.G., and J.P. received partial support from the RIT's Center for Computational Relativity and Gravitation (CCRG). 

Computational resources were provided by TACC's Frontera supercomputer allocations (Grants No. PHY20010 and No. AST-20021). Additional resources were provided by CCRG’s BlueSky and Green Prairies and Lagoon clusters acquired with NSF Grants No. PHY-2018420, No. PHY-0722703, No. PHY-1229173, and No. PHY1726215.

\section*{Appendix}
\subsection*{Photon Source Analysis}\label{sec:photon_timing}

\begin{figure}
\centering
\includegraphics[width=1.0\linewidth]{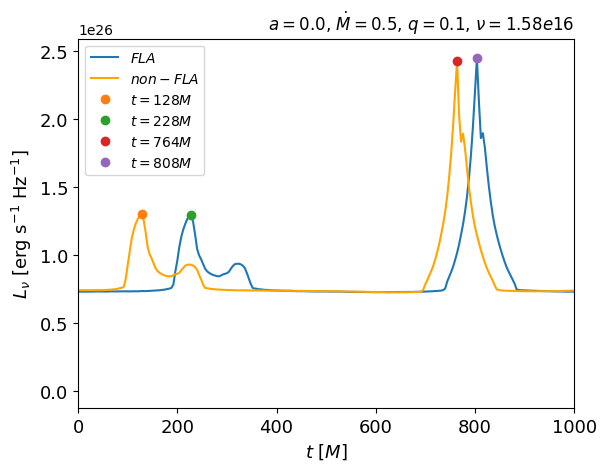}
\caption{The lightcurves for a single orbit of models M9 (FLA) and M10 (non-FLA), which both have a mass ratio of $ q = 0.1$. }\label{fig:FLAlc}
\end{figure}

\begin{figure}
\centering
\includegraphics[width=1.0\linewidth]{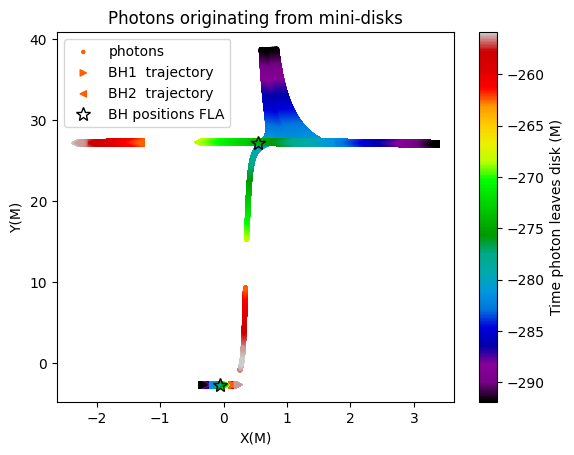}
\caption{The initial locations and times of the photons from the non-FLA models contributing to the primary flare in Figure \ref{fig:FLAlc}. These photons leave the disks between $t =-292 M $ and $t = -256M$ and arrive at the camera at $t = 764M$. The trajectories of the primary and secondary black holes are indicated by the right and left arrows respectively. The positions of the black holes at $t = 808M $ in the FLA run are indicated by black stars.}\label{fig:timing_primary}
\end{figure}

\begin{figure}
\centering
\includegraphics[width=1.0\linewidth]{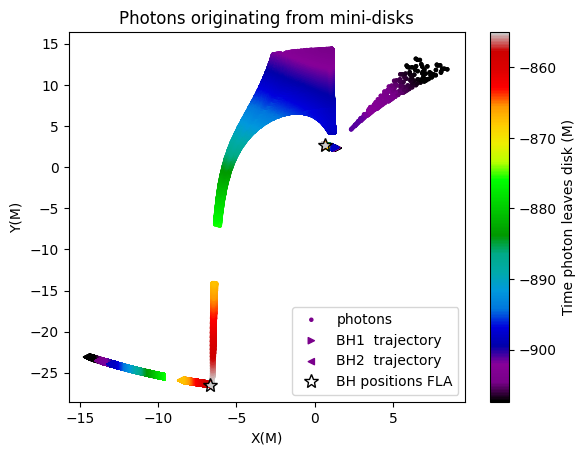}
\caption{The initial locations and times of the photons from the non-FLA models contributing to the secondary flare in Figure \ref{fig:FLAlc}. These photons leave the disks between $t =-907 M $ and $t = -855M$ and arrive at the camera at $t = 128M$. The trajectories of the primary and secondary black holes are indicated by the left and right arrows respectively. The positions of the black holes at $t = 228M $ in the FLA run are indicated by black stars.}\label{fig:timing_secondary}
\end{figure}

For the following discussion, we consider the photons that contribute to the flares exhibited in Figure \ref{fig:FLAlc}. This plot shows the light curves for the FLA (M9) and non-FLA (M10) runs for $q = 0.1$. For this figure, the light curve of the non-FLA run has not been shifted forward in time by 1000M, and the primary flares were not aligned using max overlap analysis. This is to demonstrate the differences in the arrival and departure times of the light between the FLA and non-FLA models. For the FLA model, because the speed of light is assumed to be infinitely fast, the photons leave the mini-disks at the same time they arrive at the camera, which is at $t = 228M$ for the secondary flare and $t = 808M$ for the primary flare. For the non-FLA runs the photons leave the disks at much earlier times and reach the camera at $t = 128M$ for the secondary flare and $t = 764M$ for the primary flare.

Figure \ref{fig:timing_primary} shows the origin of the photons for the non-FLA model that make up the primary flare, whereas Figure \ref{fig:timing_secondary} shows the origin of the photons contributing to the secondary flare. Both figures plot the x- and y-coordinates of the photons' emission sites from the mini-disks. Because this is a thin disk model, the z-coordinate for all photons shown in these figures is zero. The color bars give the times that the photons leave the disks. They have a range of negative values because the geodesics are integrated backward in time from the camera to their source position. Both scatter plots also show the trajectories of the black holes during the geodesic integration, with the colors corresponding to the times when the photons leave the disks. Additionally, the positions of the black holes at the time of each flare for the FLA geodesic integration have been labeled.

We can use these plots to better understand the timing difference between FLA and non-FLA runs for the secondary flares in models with mass ratios less than 1 (Figures \ref{fig:M5vM6} - \ref{fig:M9vM10}). The primary flare corresponds to the moment when the secondary black hole is located behind the primary black hole with respect to the observer. Figure \ref{fig:timing_primary} shows that during the geodesic integration, the position of the primary black hole does not change significantly, whereas, the position of the secondary black hole does. When we consider the orbital position of the black holes for both the non-FLA and FLA models at the times when the photons contributing to the primary flare are leaving the disk, we see that the majority of the photons originate from around the secondary black hole. Despite the light taking time to travel from the source to the lens, the lens' position remains relatively unchanged between the FLA and non-FLA models, during the geodesic integration.

For the secondary flare, the light from the primary black hole travels to the secondary black hole, which is moving faster and travels a greater distance during the integration. In Figure \ref{fig:timing_secondary}, we can see that the majority of the photons that contribute to the secondary flare originate from the mini-disk around the primary black hole, and this light then travels to the secondary black hole and is lensed toward the observer. Due to both the time it takes for the light to travel from the primary black hole to the secondary black hole and the faster orbital velocity of the smaller black hole, the light contributing to the secondary flare in the non-FLA model corresponds to an earlier orbital configuration of the binary system compared to the FLA model. This results in a shorter time interval between the primary and secondary flares in the non-FLA light curve.

\bibliography{references_LOCAL,biblio_LOCAL,ref}
\bibliographystyle{aasjournal}

\end{document}